\documentclass[11pt]{article}

\usepackage[pagewise]{lineno}
\usepackage[utf8]{inputenc}
\usepackage{graphicx}
\graphicspath{{figs/}}
\usepackage{subcaption}
\usepackage{amsmath, amssymb}
\numberwithin{equation}{section}
\usepackage{xcolor}
\usepackage{natbib}
\usepackage{bm}
\usepackage{nicefrac}
\usepackage{booktabs}
\usepackage{tabularx,tabulary,array}
\newcolumntype{M}[1]{>{\centering\arraybackslash}m{#1}}
\newcolumntype{N}{@{}m{0pt}@{}}
\usepackage{multirow}
\usepackage{dcolumn}
\usepackage{lscape}
\usepackage[normalem]{ulem}
\usepackage[title]{appendix}
\usepackage{authblk}
\usepackage{hyperref}
\usepackage{adjustbox}
\usepackage{booktabs}

\usepackage{multirow}
\usepackage{rotating,tabularx}

\usepackage{authblk}
\usepackage[english]{babel}
\usepackage[T1]{fontenc}

\usepackage{amssymb, amsmath, amsthm, blkarray}
\usepackage{graphicx, tikz}
\usepackage{natbib}

\definecolor{mylightyellow}{rgb}{1,1,.8}
\definecolor{mylightgreen}{rgb}{.8,1,.8}
\definecolor{mydarkred}{RGB}{178,34,34}
\definecolor{mydarkgreen}{RGB}{34,139,34}
\definecolor{mydarkblue}{RGB}{72,61,139}
\definecolor{mydarkyellow}{RGB}{218,165,32}

\usepackage{booktabs}

\usepackage{marginnote}
\usepackage{soul}

\usepackage[a4paper,margin=15mm]{geometry}

\usepackage[]{hyperref}
 \hypersetup{
 colorlinks=true,breaklinks=true,
 urlcolor= mydarkblue,linkcolor= mydarkblue,citecolor= mydarkgreen,
 pdfauthor={S\'aez de Oc\'ariz Borde, H., Sondak, D., Protopapas, P.}
}

\theoremstyle{plain}
\theoremstyle{definition}

\usepackage{eurosym}

\title{Multi-Task Learning based Convolutional Models with Curriculum Learning for the Anisotropic Reynolds Stress Tensor in Turbulent Duct Flow}

\author[1,2]{Haitz S\'aez de Oc\'ariz Borde\thanks{\tt haitz.saez-de-ocariz-borde17@imperial.ac.uk}}
\author[3,1]{David Sondak\thanks{\tt david.sondak@3ds.com}}
\author[1]{Pavlos Protopapas\thanks{\tt pavlos@seas.harvard.edu}}

\affil[1]{\small Institute for Applied Computational Science, Harvard University, Cambridge, MA 02138,
United States}
\affil[2]{\small Department of Aeronautics, Imperial College London, London SW7 2AZ, United Kingdom}
\affil[3]{\small Dassault Systemes, Simulia Corp.}

\date{
\small First Version: May 11, 2021.  This version: \today
}

\begin{document}

\maketitle

\begin{abstract}
The Reynolds-averaged Navier-Stokes (RANS) equations require accurate modeling of the anisotropic Reynolds stress tensor. Traditional closure models, while sophisticated, often only apply to restricted flow configurations. Researchers have started using machine learning approaches to tackle this problem by developing more general closure models informed by data. In this work we build upon recent convolutional neural network architectures used for turbulence modeling and propose a multi-task learning-based fully convolutional neural network that is able to accurately predict the normalized anisotropic Reynolds stress tensor for turbulent duct flows. Furthermore, we also explore the application of curriculum learning to data-driven turbulence modeling.
\end{abstract}

\bigskip

\noindent {\bf Keywords:} turbulence modeling, Reynolds-averaged Navier-Stokes, duct flow, convolutional neural networks, multi-task learning, curriculum learning.




\newpage
\section{Introduction}
\label{sec:introduction}

The equations governing fluid mechanics have long been established, but fluid turbulence still remains one of the greatest unsolved problems of classical physics. Scientists and engineers seek to develop turbulence models with good performance that are tractable from a computational perspective and useful for real-world applications. Direct numerical simulations (DNS) of turbulent fluid systems are considered the gold standard because they do not require any additional models; the flow physics is fully resolved up to the available computational resources. In this approach, the Navier-Stokes equations are discretized using well-refined grids so that all scales of motion can be resolved. Using this technique, researchers have managed to obtain mathematical correlations that would otherwise be intractable if they were to rely solely on experimental measurements~(\cite{dns_review}). Nevertheless, even according to optimistic estimates, the computing power required for DNS simulations on the vast majority of practical flows may not become available until the end of this century~(\cite{dns_couette}). Given that most problems remain intractable using DNS simulations, researchers have developed approaches such as Reynolds-averaged Navier-Stokes (RANS) models and Large Eddy Simulation (LES) models~\cite{alma993553114401591}. RANS models are less accurate than LES models, but they are also less computationally expensive and their use is more widespread in industrial applications. The objective of RANS models is to find the average velocity and pressure fields, but this approach induces the Reynolds stress tensor which encompasses the effects of the fluctuating velocity field on the average fields. Modeling efforts in RANS are focused on developing suitable models for the Reynolds stress tensor.

Considering that the applicability of traditional two-equation eddy viscosity models such as the $k-\epsilon$ and the $k-\omega$ is limited depending on the flow (see~\cite{speziale,johansson2002engineering, chen2003extended, gatski2004constitutive}), researchers have started working on applying machine learning (ML) and artificial intelligence (AI) based algorithms to develop closure models for RANS that can potentially work across different flows, including those for which applying traditional models may be troublesome (see~\cite{ling_ml,ling2016reynolds,fang2018deep,PhysRevFluids.3.074602,Kaandorp2018MachineLF,song,KAANDORP2020104497,zhu2020datadriven}). \cite{ling2016reynolds} proposed the Tensor Basis Neural Network (TBNN) to learn the coefficients of an integrity basis for the Reynolds anisotropy tensor from~\cite{pope_1975}. Inspired by the TBNN, \cite{fang2018deep} used a fully-connected neural network to model the Reynolds stress for fully-developed turbulent channel flow and incorporated different physics embedding techniques to the baseline model, such as friction Reynolds number injection and boundary condition enforcement, amongst others. \cite{borde2021convolutional} further built on the work by~\cite{fang2018deep} and proposed a better-performing baseline convolutional neural network (CNN) model which also incorporated physics embedding techniques and tested its applicability to a number of one-dimensional turbulent flows. Additionally, \cite{borde2021convolutional} discussed several interpretability techniques to drive the model design and provide guidance on the model behavior in relation to the underlying physics of the problem. Other researchers such as~\cite{jiang} have also explored machine learning model interpretability in a turbulence modeling context. A survey on data-driven turbulence modeling can be found in~\cite{duraisamy}. 

In this paper, we extend previous work on CNNs by~\cite{borde2021convolutional} beyond simple one-dimensional turbulent flows, by investigating the applicability of such models and physics embedding techniques to turbulent duct flow. We propose a novel deep CNN architecture with several ramifications based on physics-informed multi-task learning (MTL) (\cite{multi-task_learning,mt_thung,zhang2021surveymt}). The architecture is able to efficiently capture the input-output relationship in the data for a wide range of bulk Reynolds numbers using a relatively low number of trainable network parameters. In particular, we use a hard parameter sharing MTL approach in which common layers are shared upstream of the model. Downstream, the model branches out into specialized layers for each task, i.e., for predicting different components of the anisotropic Reynolds stress tensor. Apart from helping to drastically reduce the number of model parameters required to train the network, MTL improves model attention, regularization, and model generalizability. 


We also explore curriculum learning for turbulence modeling. Curriculum learning is a training strategy that trains a ML model starting from easier to harder examples, which mimics the learning order in human curricula~(\cite{ELMAN199371,rohde}). Several curriculum learning strategies have already demonstrated their power in improving convergence and generalization capabilities of a number of models in tasks such as natural language processing and computer vision~(\cite{cl_survey}).

The rest of the paper has the following structure. Section~\ref{sec:Background} covers background information on the RANS equations, turbulent duct flow, the dataset used in this work, neural networks, MTL and curriculum learning. Section~\ref{sec:Convolutional Neural Network Model for Turbulent Duct Flow} reviews the new CNN model proposed for turbulent duct flow. Section~\ref{sec:Results} covers the performance of the CNN model for turbulent duct flow and compares training with and without curriculum learning. Lastly, Section~\ref{sec:Conclusion} summarizes the final conclusions and gives hints for future work.

\section{Background and Methodology}
\label{sec:Background}

\subsection{The RANS Equations}
\label{subsec:The RANS equations}

The RANS equations are derived from the Navier-Stokes by introducing a sum decomposition the velocity and pressure fields into average components, $\mathbf{u}=\overline{\mathbf{u}}+\mathbf{u^{\prime} }$ and $p=\overline{p}+p^{\prime}$, and then averaging the Navier-Stokes equations. The aim is to find the average velocity  $\overline{\mathbf{u}}$ and pressure $\overline{p}$. Applying this averaging operation to the Navier-Stokes equations we obtain the RANS equations. Compared to the Navier-Stokes equations, the RANS equations include an additional term $\overline{\mathbf{u^{\prime}}\otimes\mathbf{u^{\prime}}}$ known as the Reynolds stress tensor. The Reynolds stress tensor must be modeled to close the RANS equations. This modeling effort has been the subject of intense research for several decades. Researchers are mainly focused on modeling the anisotropic Reynolds stress tensor $\mathbf{a}=\overline{\mathbf{u^{\prime}}\otimes\mathbf{u^{\prime}}}-(2k/3)\mathbf{I}$, because that is the portion responsible for turbulent transport, where $k$ is the turbulent kinetic energy $k=\frac{1}{2}\text{trace} (\overline{\mathbf{u^{\prime}}\otimes\mathbf{u^{\prime}}})$. The neural network models in this work are trained based on the normalized anisotropy tensor, $\mathbf{b} = \mathbf{a}/(2k)$. We denote individual components of the tensor with subscripts referring to the fluctuating velocity components involved (e.g. $b_{uw} = \overline{u^{\prime}w^{\prime}} / (2k)$).

\subsection{Turbulent Square Duct Flow}
\label{subsec:Turbulent Duct flow}
Turbulent duct flow is an interesting problem from both the engineering and the academic perspectives. Ducts are often used as passages in ventilation systems, heating, and air conditioning to both remove and deliver air. The accurate prediction of turbulent flow in ducts is important in many practical applications such as in the aerospace~(\cite{duct_aerospace}), agrofood~(\cite{agrofood}), process~(\cite{process}), and nuclear~(\cite{nuclear}) industries to name a few. From an academic standpoint, the flow in a square duct is a challenging test case for most EVMs, since they are unable to predict the secondary corner vortices that form in the duct~(\cite{secondary}). Furthermore, even the Reynolds-stress transport model (RSTM) can face complications when trying to predict this behaviour well~(\cite{Gerolymos_2015,gerorms,sormc,3dre}). In particular, for turbulent fully-developed flow in a straight duct~(\cite{gessner_jones_1965}) the anisotropy of the diagonal stresses in the crossflow plane, $a_{vv}$ and $a_{ww}$~(\cite{peter}), and the inhomogeneity of the gradients of the secondary shear-stress, $a_{vw}$~(\cite{brundrett_baines_1964}), trigger the secondary flow associated with streamwise vorticity~(\cite{Gerolymos_2015}). Previous work by~\cite{wu_again} and ~\cite{laizet} has suggested that important physical information may be lost if some key invariants are omitted as inputs to data-driven turbulence ML models and that including quantities such as the pressure gradients may be beneficial and help overcome some of the prediction problems encountered by EVMs.

To train and test the model we use the dataset by~\cite{mcconkey2021curated} accessible at Kaggle\footnote{https://www.kaggle.com/ryleymcconkey/ml-turbulence-dataset.}. Various RANS features with DNS/LES labels are available in the dataset for different flow configurations. The flow is assumed to be incompressible, viscous, statistically steady state, and turbulent for all cases. We focus on the square duct data which is based on the work by~\cite{pinelli_uhlmann_sekimoto_kawahara_2010}. The data considers a range of bulk Reynolds numbers from a marginal state up to fully-developed turbulent flow at low bulk Reynolds numbers ranging between $Re_{b}=1100$ and $Re_{b}=3500$. The bulk Reynolds number is defined as $Re_{b} = \frac{U_{b}h}{\nu}$ and consists of the bulk velocity $U_{b}$, the duct half-width $h$ and the
kinematic viscosity $\nu$. As suggested by~\cite{marginal}, DNS simulations seem to establish that in the case of a straight square duct self-sustaining turbulence can be maintained for bulk Reynolds numbers above $Re_{b}=1100$. We will use the input features from the $k-\epsilon-\phi_{t}-f$ model in the dataset as recommended by~\cite{mcconkey2021curated}. The $k-\epsilon-\phi_{t}-f$ model is an improved version of both the original $\overline{v^{\prime}}^{2}-f$ model developed by~\cite{Durbin1991NearwallTC} and the enhanced version by~\cite{lien_version}. In their simulation for the square duct, \cite{mcconkey2021curated} use a duct of dimensions $2h \times 2h \times 5h$ and apply wall boundary conditions for the top, bottom and sides of the
duct, so as to match the DNS by~\cite{pinelli_uhlmann_sekimoto_kawahara_2010}. For more information on the input features and labels the reader is referred to~\cite{mcconkey2021curated}.

The RANS equations for fully-developed steady duct flow are,
\begin{align}
\overline{v}\frac{\partial \overline{u}}{\partial y}+\overline{w}\frac{\partial \overline{u}}{\partial z} &= -\frac{1}{\rho}\frac{d\overline{p}}{dx}+\nu\nabla^{2}\overline{u}-\frac{\partial\big( \overline{u^{\prime}v^{\prime}}\big)}{\partial y}-\frac{\partial\big( \overline{u^{\prime}w^{\prime}}\big)}{\partial z}  \label{eq:RANSduct1} \\
\overline{v}\frac{\partial \overline{v}}{\partial y}+\overline{w}\frac{\partial \overline{v}}{\partial z} &= -\frac{1}{\rho}\frac{d\overline{p}}{dy}+\nu\nabla^{2}\overline{v}-\frac{\partial\big( \overline{v^{\prime}v^{\prime}}\big)}{\partial y}-\frac{\partial\big( \overline{v^{\prime}w^{\prime}}\big)}{\partial z} \label{eq:RANSduct2} \\
\overline{v}\frac{\partial \overline{w}}{\partial y}+\overline{w}\frac{\partial \overline{w}}{\partial z} &= -\frac{1}{\rho}\frac{d\overline{p}}{dz}+\nu\nabla^{2}\overline{w}-\frac{\partial\big( \overline{v^{\prime}w^{\prime}}\big)}{\partial y}-\frac{\partial\big( \overline{w^{\prime}w^{\prime}}\big)}{\partial z} \label{eq:RANSduct3} \\
\frac{\partial \overline{v}}{\partial y}+\frac{\partial \overline{w}}{\partial z} &= 0 \label{eq:RANSduct4}
\end{align}
We assume that the componentes of the anisotropic Reynolds stress tensor can generally be represented as in the functional form
\begin{equation}
    \big(a_{uv},a_{uw},a_{vv},a_{vw},a_{ww}\big)=f\left(\frac{\partial \overline{u}}{\partial y},\frac{\partial \overline{u}}{\partial z},\frac{\partial \overline{v}}{\partial y},\frac{\partial \overline{v}}{\partial z},\frac{\partial \overline{w}}{\partial y},\frac{\partial\overline{p}}{\partial x},\frac{\partial\overline{p}}{\partial y},\frac{\partial\overline{p}}{\partial z},Re_{b}\right)\,.
    \label{eq:input and output}
\end{equation}

Given that the equations are coupled and the exact relation between input and output is unknown, we predict all relevant components of the anisotropic Reynolds stress tensor based on all key invariant inputs as suggested by~\cite{wu_again} and~\cite{laizet}. \eqref{eq:input and output} will be normalized later in Section~\ref{sec:Convolutional Neural Network Model for Turbulent Duct Flow}. Note that the mean pressure $\overline{p}$ is a zeroth-order tensor, that is, a scalar physical quantity, which is invariant under any coordinate transformation~(\cite{craft}) including the Galilean transformation, and the pressure gradient is also equivalent in any two frames of reference.

\subsection{Neural Networks}
\label{subsec:Neural Networks}
Deep Feedforward Networks, also called Feedforward Neural Networks or Multilayer Perceptrons (MLPs), are considered the archetype of Deep Learning (DL) models. Feedforward networks are used to obtain an approximation of some function $f$. For instance, $\mathbf{y}=f(\mathbf{x})$ maps an input $\mathbf{x}$ to a prediction $\mathbf{y}$. A feedforward network defines a mapping $\mathbf{y}=f^{*}(\mathbf{x};\boldsymbol{\theta})$ and is optimized to find the parameters $\boldsymbol{\theta}$ that give the best approximation of the function $f$. In feedforward models the information flows through the function being evaluated from $\mathbf{x}$, through the intermediate computations and finally to the output $\mathbf{y}$, as described by~\cite{Goodfellow-et-al-2016}. 

In the case of a Fully-Connected Feedforward Neural Network (FCFF), the output of a given layer $i$ is passed to layer $i+1$ and every node in the current layer is connected to every node in the previous and
subsequent layer, without any jump connections or feedback loops. On the other hand, CNNs use convolutional layers where each neuron is only connected to a few nearby neurons in the previous layer. In the case of convolutional layers, the same set of weights are shared by every neuron. This connection pattern is applied to cases where the data can be interpreted as being spatial. The convolutional layer's parameters consist of a set of learnable filters or kernels. Every filter has relatively small dimensions, but extends through the full depth of the input volume. A convolution is computed by sliding the filter over the input or the output of the previous layer. At every location, a matrix multiplication is performed and sums the result onto the feature map. The feature map is the output of the filter applied to the previous layer. In general, CNNs may also have pooling layers and fully-connected layers before the output. See~\cite{millstein2018convolutional} for an introduction to CNNs. 

For both FCFFs and CNNs, the model predictions are compared to data in a loss function, and an optimization algorithm (see \cite{ml_opt}), such as stochastic gradient descent (see \cite{sgd} and \cite{DBLP:journals/corr/Ruder16}), is used to adjust all the weights and biases, $\boldsymbol{\theta}$, of the network to minimize the loss function. The process of optimizing a neural network can be challenging since it is a high-dimensional non-convex optimization problem.

Adopting neural networks for physics-based problems entails the challenge of trying to inform the network with known physical laws, since otherwise the network will simply try to memorize the patterns in the data without paying attention to the underlying physics. One of the first attempts to embed the physical and mathematical structure of the Reynolds anisotropy tensor into a neural network was the TBNN by~\cite{ling2016reynolds}. The TBNN guaranteed Galilean and rotational invariance of the predicted Reynolds anisotropy tensor by adding an additional tensorial layer to a FCFF network which returned the most general, local eddy viscosity model described in~\cite{pope_1975}. \cite{fang2018deep} further proposed a number of techniques such as reparameterizing a FCFF network to enforce the no-slip boundary condition as a function of the normalized distance from the wall, explicitly providing $Re_{\tau}$ to the network to incorporate friction Reynolds number information into the model and extensions to allow for non-locality. Later, \cite{borde2021convolutional} extended the physics embedding techniques proposed by~\cite{fang2018deep} to CNNs.

\subsection{Multi-Task Learning}
\label{subsec:Multi-Task Learning}

Multi-task learning (MTL) consists in optimizing multiple related learning tasks at the same time by leveraging their shared information to enhance generalization and the model performance for each task (\cite{multi-task_learning,mt_thung,zhang2021surveymt}). It is closely related to other ML subfields such as multi-class learning (\cite{Aly2005SurveyOM}) and transfer learning (\cite{5288526}) and it has successfully been implemented across a wide range of ML applications, namely, computer vision (\cite{fast_r_cnn}), natural language processing (\cite{Collobert2008AUA}), speech recognition (\cite{Deng2013NewTO}), drug discovery (\cite{ramsundar2015massively}) and stock selection (\cite{stock_mt}). 
MTL learning can be classified into hard parameter sharing MTL and soft parameter sharing MTL. Hard parameter sharing is the most commonly used MTL approach and consists in sharing the hidden layers between all tasks while reserving some output-specific layers downstream of the model for each task. It has been shown that this approach can substantially reduce overfitting. Indeed, the risk of overfitting the shared parameters is reduced by an order $N$, where $N$ is the number of tasks, as compared to the parameters in the output-specific layers, as discussed by~\cite{Baxter97abayesian/information}. On the other hand, in soft parameter sharing MTL, multiple neural network models are used for each task and a regularization method is implemented to encourage the trainable parameters of each model to be similar (\cite{Duong2015LowRD,Yang2017TraceNR}). Note that in this approach the models do not directly share parameters. In this work we will implement hard parameter sharing MTL for the convolutional model for turbulent square duct flow, since we are interested in finding an efficient model that is fast to train and which has as few parameters as possible.

There are several underlying mechanisms that make MTL work. MTL is closely related to implicit data augmentation, since in practical terms, it increases the sample size we are using for training the model. MTL also improves attention focusing of the model and it can help the network better identify which features really matter to  successfully learn the overall input-output relationship in the data. Neural networks may find it difficult to ignore data-dependent noise, especially when data is limited and high-dimensional. Using MTL and trying to optimize for different tasks will provide additional evidence for the relevance or irrelevance of features. At the same time, MTL may help identify features and relationships in the input data that may be important for all tasks, but easier to identify when optimizing for a particular objective. This is referred to as ``eavesdropping''. Lastly, given that MTL makes the model learn representations that all tasks prefer, it helps with regularization and makes the model more likely to better generalize to new tasks from the same environment, as discussed by~\cite{bias_learning}. Note, however, that some requisites must be met to make MTL work. Firstly, the capacity of the shared layer modules must be enough to capture the complexity of the problem and not to cause interference between tasks. At the same time, if the capacity was too large, it could hinder transfer of knowledge, since the model could simply learn independent features for each task. Also, the covariance between tasks must be taken into consideration, that is, the tasks should be related to one another. Task covariance is a measure of the alignment of the input data for different tasks among their principal directions. The intuition behind task covariance is that if the principal directions of the data for different tasks is not well-aligned, feeding the data into the shared module can result in suboptimal models rather than help learning (\cite{senwu}). 

\subsection{Curriculum Learning}
\label{subsec:Curriculum Learning}
Curriculum learning was proposed during the 1990s by researchers that worked at the intersection of cognitive science and ML such as~\cite{ELMAN199371} and~\cite{rohde}. Curriculum learning is the idea that ML algorithms may benefit from progressively learning from simple concepts to hard problems in the same way that humans do. The work by~\cite{ELMAN199371} addressed the possible cooperative interactions between maturement and the capacity to learn a complex domain, and suggested that instead of being a limitation, restrictions on resources during development may be a prerequisite for successful learning and mastering of some complicated domains. Later, \cite{cl_bengio} showed that significant improvements in generalization could be obtained using this technique in the context of training using non-convex criteria for deep deterministic and stochastic neural networks. They further hypothesized that curriculum learning could both positively affect the speed of convergence of the training process to a minimum and, at the same time, in the case of non-convex criteria improve the quality of the local minima achieved.

Curriculum learning may be regarded as a continuation method (\cite{continuation}); a general strategy for global optimization of non-convex functions. Despite such methods not guaranteeing ultimate convergence to the global minimum, starting the optimization using a smoothed version of the objective function is more likely to unveil the global picture of the problem. Hence, the intuition behind curriculum learning is to optimize a smoothed objective function and gradually consider less
smoothing by feeding more complicated examples to the algorithm. As a basic example, we may start by defining a family of losses, $L_{\lambda}(\boldsymbol{\theta})$, where $L_{0}(\boldsymbol{\theta})$ can be optimized fairly easily, say $L_{0}(\boldsymbol{\theta})$ is convex in $\boldsymbol{\theta}$. On the other hand, we can define $L_{1}(\boldsymbol{\theta})$ as the criterion of interest, which is not smooth and is difficult to optimize. $L_{0}(\boldsymbol{\theta})$ being a substantially smoothed version of $L_{1}(\boldsymbol{\theta})$, we aim at minimizing $L_{0}(\boldsymbol{\theta})$ so as to move $\boldsymbol{\theta}$ to the basin of attraction of a dominant minimum of the criterion of interest, which may be the global minimum. Then we gradually increase $\lambda$ while keeping $\boldsymbol{\theta}$ at a minimum of $L_{\lambda}(\boldsymbol{\theta})$. This is likely to be advantageous as compared to starting the optimization problem for $L_{1}(\boldsymbol{\theta})$ using initial random values for $\boldsymbol{\theta}$, which may lead to getting stuck in a poor local minimum.






To the best of our knowledge, curriculum learning has not yet been explored in the context of data-driven turbulence modeling. Curriculum learning has the potential to speed up the training process of complex data-driven ML models that may have a substantial computational cost associated with them, especially for complex high-dimensional flows. We will apply curriculum learning to try to improve the performance of our CNN model for turbulent square duct flow.

\section{Convolutional Neural Network Model for Turbulent Square Duct Flow}
\label{sec:Convolutional Neural Network Model for Turbulent Duct Flow}

In this section we present a new two-dimensional fully-convolutional architecture for turbulent square duct flow based on MTL. This network, which we will refer to as the Multi-Task Learning Convolutional Neural Network (MTLCNN), is an extension of the CNN models in~\cite{borde2021convolutional} to higher-dimensional flows. 
The MTLCNN model uses the non-dimensional velocity gradients
\begin{equation}
\left(\frac{\partial \overline{u}}{\partial y}^{*},\frac{\partial \overline{u}}{\partial z}^{*},\frac{\partial \overline{v}}{\partial y}^{*},\frac{\partial \overline{v}}{\partial z}^{*},\frac{\partial \overline{w}}{\partial y}^{*}\right) = \frac{k}{\epsilon}\left(\frac{\partial \overline{u}}{\partial y},\frac{\partial \overline{u}}{\partial z},\frac{\partial \overline{v}}{\partial y},\frac{\partial \overline{v}}{\partial z},\frac{\partial \overline{w}}{\partial y}\right),
\end{equation}
and the non-dimensional pressure gradient,
\begin{equation}
    \left(\frac{\partial\overline{p}}{\partial x}^{*},\frac{\partial\overline{p}}{\partial y}^{*},\frac{\partial\overline{p}}{\partial z}^{*}\right)=\dfrac{1}{\left\|\nabla k\right\|_{2}}\left(\frac{\partial\overline{p}}{\partial x},\frac{\partial \overline{p}}{\partial y},\frac{\partial\overline{p}}{\partial z}\right),
\end{equation}
where $\epsilon$ is the turbulent dissipation rate, $\left\|\nabla k \right\|_{2}$ is the Euclidean norm of the gradient of the turbulent kinetic energy. It also uses the normalized Reynolds anisotropy tensor, $\mathbf{b}=\mathbf{a}/(2k)$, and the bulk Reynolds number, $Re_{b} = U_{b}h/\nu$, where $U_{b}$ is the bulk velocity, $h$ is the duct half-width and $\nu$ the kinematic viscosity.

The model takes as input the normalized mean velocity gradients, the normalized mean pressure gradients and the bulk Reynolds number~(\cite{fang2018deep}), and aims at predicting the normalized anisotropic Reynolds stress tensor. The input-output relationship is represented as
\begin{equation}
    \left(b_{uv},b_{uw},b_{vv},b_{vw},b_{ww}\right)=\text{MTLCNN}\left(\frac{\partial \overline{u}}{\partial y}^{*},\frac{\partial \overline{u}}{\partial z}^{*},\frac{\partial \overline{v}}{\partial y}^{*},\frac{\partial \overline{v}}{\partial z}^{*},\frac{\partial \overline{w}}{\partial y}^{*},\frac{\partial\overline{p}}{\partial x}^{*},\frac{\partial\overline{p}}{\partial y}^{*},\frac{\partial\overline{p}}{\partial z}^{*},Re_{b}\right).
    \label{eq:CNN_eq}
\end{equation}

Note that the $Re_{b}$ information is injected into the model by concatenating to the input a matrix of constant value equal to the bulk Reynolds number. Given that the equations for turbulent duct flow are coupled, it is sensible to start from the assumption that all components of the stress tensor may depend on all the input velocity gradients, the pressure gradients and the bulk Reynolds number and to let the algorithm learn the unknown relationship between the input and the output. The complexity of the problem has substantially increased compared to one-dimensional flows. The MTLCNN will take as input for each bulk Reynolds number case a tensor with 9 channels, one for each parameter in $\left(\frac{\partial \overline{u}}{\partial y}^{*},\frac{\partial \overline{u}}{\partial z}^{*},\frac{\partial \overline{v}}{\partial y}^{*},\frac{\partial \overline{v}}{\partial z}^{*},\frac{\partial \overline{w}}{\partial y}^{*},\frac{\partial\overline{p}}{\partial x}^{*},\frac{\partial\overline{p}}{\partial y}^{*},\frac{\partial\overline{p}}{\partial z}^{*},Re_{b}\right)$ and the output of the MTLCNN will have 5 channels one for each entry of the normalized anisotropic Reynolds stress tensor in $\left(b_{uv},b_{uw},b_{vv},b_{vw},b_{ww}\right)$. The proposed MTLCNN model has multiple layers with multiple filters and it is fully-convolutional.

\subsection{Data Pre-processing}
\label{Data pre-processing}

Although CNNs have achieved extraordinary results in many pattern recognition and standard computer vision applications, their adoption in the computer graphics and geometry processing communities is limited by the non-Euclidean and unstructured nature of the data, as discussed by~\cite{NIPS2016_228499b5}. Indeed, the same issue applies to many turbulence modeling problems in which complex meshes are employed to discretize the flow. Being able to apply CNNs to problems defined by meshes of varying resolution and shapes is a challenging problem in deep learning, especially when considering 3D geometric domains (\cite{MasciBBV15,monti,Poulenard2018MultidirectionalGN}).

\begin{figure}[htbp!]
\centering
\includegraphics[width=0.4\linewidth]{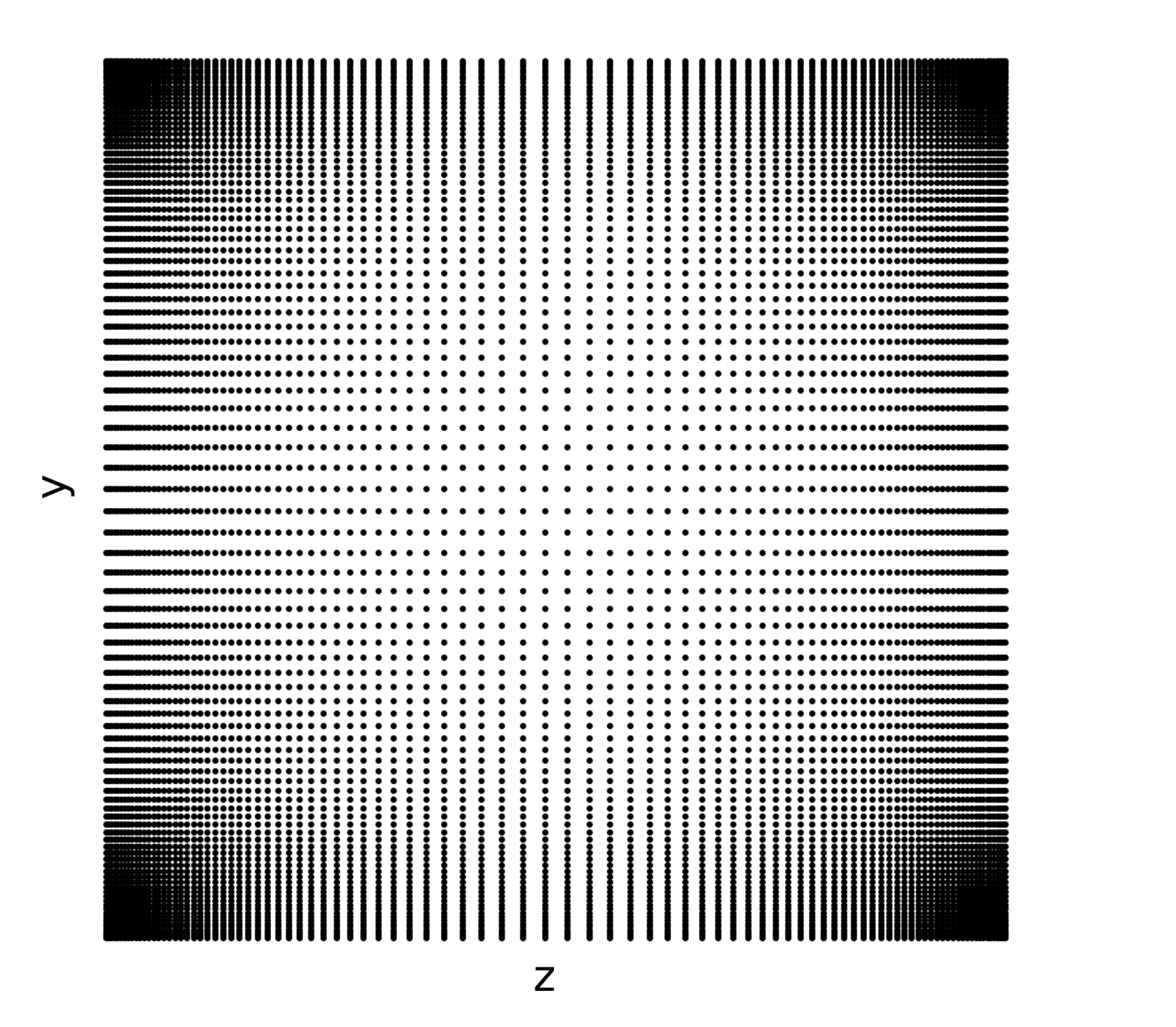}
\caption{The grid points for the square duct flow highlighting the structured non-uniform mesh in the provided dataset.}
\label{fig:duct_mesh}
\end{figure}

In the case of the square duct flow at hand, we are working with cross-sectional data. The data by~\cite{mcconkey2021curated} is computed on a structured non-uniform grid. It is provided in the form $(C_{x},C_{y},C_{z},\psi)$, where $C_{x}, C_{y}$, and $C_{z}$ are one-dimensional arrays that correspond to the $x,y$, and $z$ coordinates and the last entry $\psi$, represents the value of the parameter field at a given location. For all the cases in the dataset, there are multiple points with the same wall distance. For the square duct specifically, this arises due to symmetry of the case itself, as well as the rectilinear grid being aligned with the geometry. Figure~\ref{fig:duct_mesh} shows the cross-sectional location of the data grid points. 

\begin{figure}[htbp!]
\centering
\subfloat[]{\label{fig:full}\includegraphics[height=0.35\linewidth]{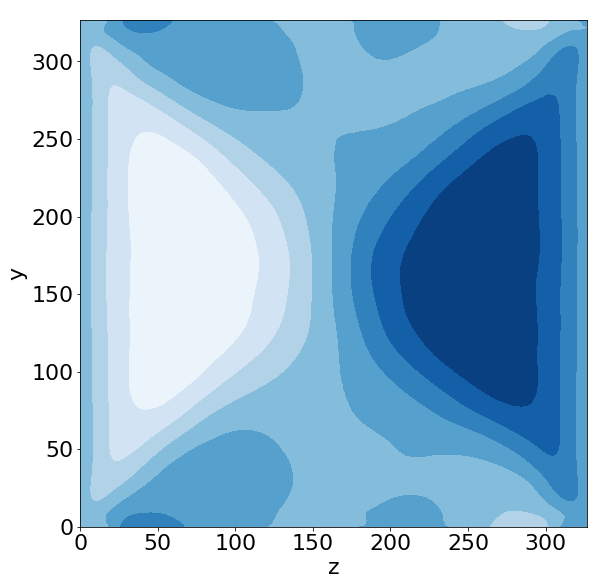}}
\subfloat[]{\label{fig:zoomin}\includegraphics[height=0.35\linewidth]{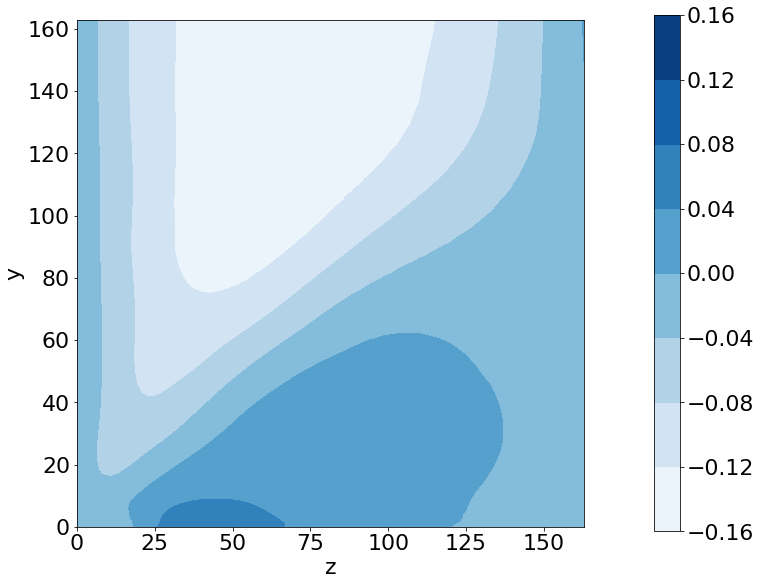}}
\caption[Contour plots of $b_{uv}$ for $Re_{b}=2400$ from the dataset. (a) Matrix containing information for the whole cross-section of the square duct ($328\times328$ matrix). (b) Matrix with data for the bottom left quadrant of the cross-section ($164\times164$ matrix).]{Contour plots of $b_{uv}$ for $Re_{b}=2400$. (a) Matrix containing information for the whole cross-section of the square duct ($328\times328$ matrix). (b) Matrix with data for the bottom left quadrant of the cross-section ($164\times164$ matrix). See the main text for a discussion.}
\label{fig:boost_comparison}
\end{figure}

The data must be pre-processed before feeding it to the MTLCNN model, so as to make it possible to apply two-dimensional convolutional operations that capture meaningful spatial patterns. Before allowing the neural network to process the data, the non-uniform data from the dataset is interpolated onto a uniform mesh
\begin{align}
  z_{\kappa} = y_{\kappa} =\left[ d_{1} + \kappa\left( \frac{d_{2}-d_{1}}{n-1}\right)\right],\,\kappa=0,1,...,n-1
\end{align}
where $n=328$, $d_{1}=-0.498505$ and $d_{2}=0.498505$ (same bounds as in the original data), by fitting a function of the form $\psi^{*} = \Gamma(C_{z},C_{y})$ to the scattered data on the cross section $(C_{z},C_{y},\psi)$ using triangulation-based linear interpolation (\cite{interpolation}). This way we obtain two-dimensional matrices for each input and label. We map the data onto matrices of dimensions $328 \times 328$. This number of entries is the minimum required to guarantee that we are effectively capturing the original resolution of the data at the corners of the duct and that no information is being lost. We calculate it considering the minimum increment used by the original mesh. Note that according to~\cite{mcconkey2021curated} the data points for the boundary values are only for the internal cells of the mesh. This means that the data provided  by~\cite{mcconkey2021curated} does not exactly reach the wall of the duct. Although in practice, especially for RANS, it is very challenging to obtain a perfectly symmetric field, we can consider one quarter of the domain in the cross-sectional plane for calculations due to the theoretical symmetry of the flow (\cite{ducts_vinuesa}). Hence, we use the bottom left quadrant of the duct flow data for calculations, as suggested by previous similar work (see~\cite{wu_again,laizet}). This means that ultimately a $164 \times 164$ matrix is fed to the MTLCNN model for each interpolated feature $\psi^{*}$, both in the case of model inputs and outputs. Figure~\ref{fig:full} and Figure~\ref{fig:zoomin} show a comparison between the data for the whole cross-sectional area and the data extracted for the lower left quadrant alone. Note that the data by~\cite{mcconkey2021curated} does not perfectly match the boundary condition at the wall in some regions (this is not due to the interpolation method used in this paper). \cite{mcconkey2021curated} attributes this to the fact that the boundary values are only for the internal cells of the mesh. This is in general not a big issue for dimensional quantities but in the case of the normalized anisotropic Reynolds stress tensor we can expect some abrupt changes near the wall in Figure~\ref{fig:boost_comparison}.


\subsection{Label Transformations}
\label{subsec: Label Transformations}
\begin{figure}[!ht]
\centering
\subfloat[]{\label{fig:b_uv2400}\includegraphics[width=0.2\linewidth]{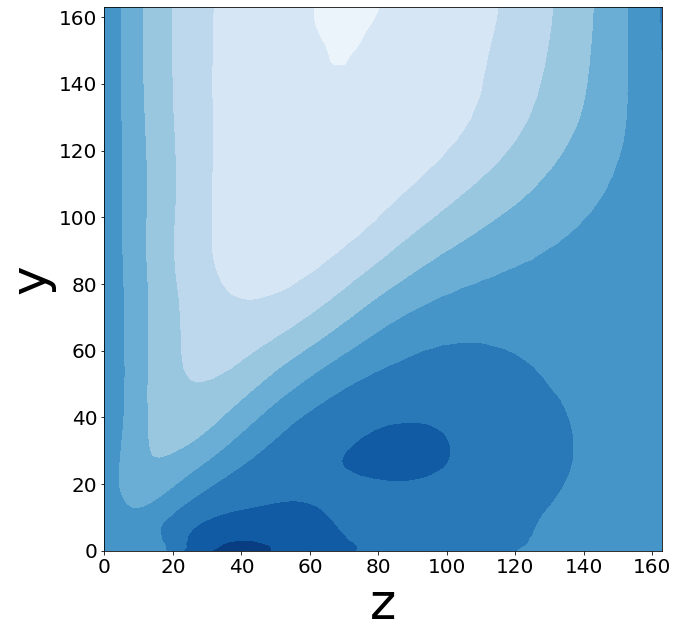}}
\subfloat[]{\label{fig:b_uw2400}\includegraphics[width=0.2\linewidth]{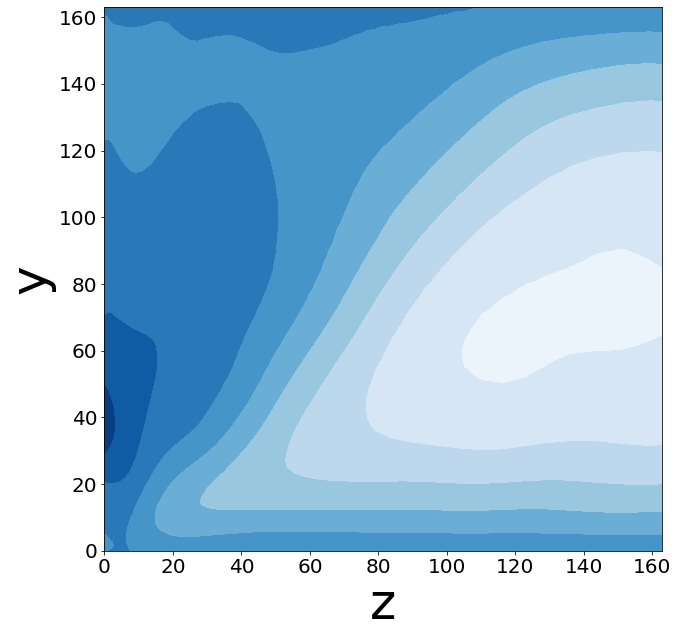}}
\subfloat[]{\label{fig:b_vv2400}\includegraphics[width=0.2\linewidth]{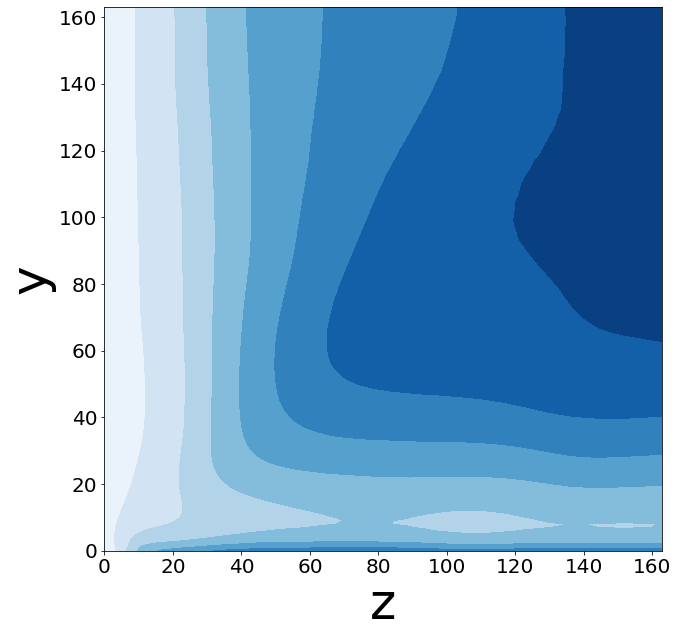}}
\subfloat[]{\label{fig:b_vw2400}\includegraphics[width=0.2\linewidth]{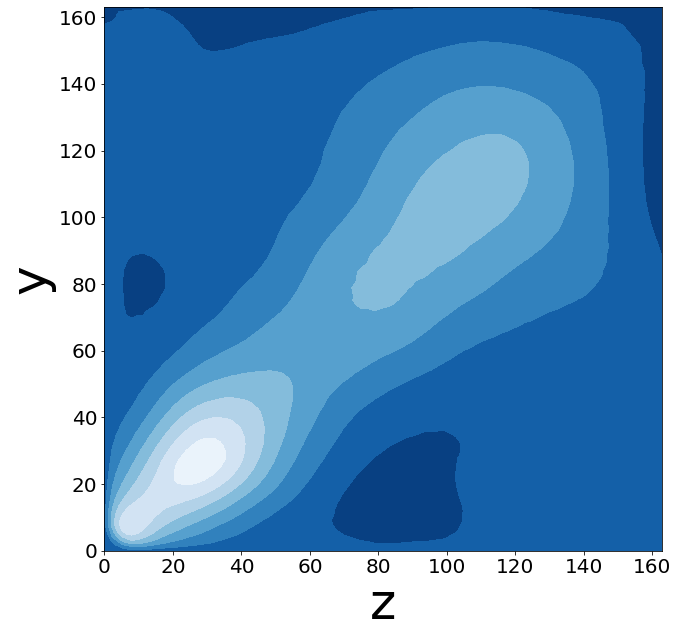}}
\subfloat[]{\label{fig:b_ww2400}\includegraphics[width=0.2\linewidth]{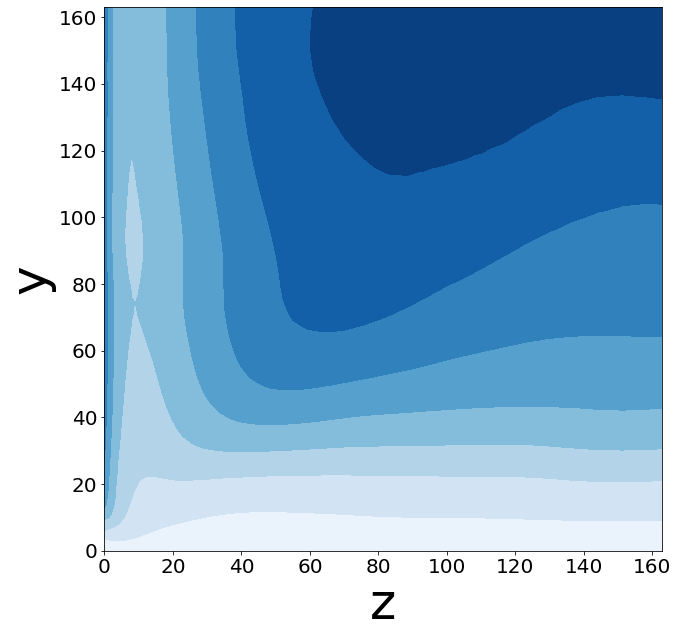}}
\caption{Original label data. Qualitative contour plots of (a) $b_{uv}$, (b) $b_{uw}$, (c) $b_{vv}$, (d) $b_{vw}$ and (e) $b_{ww}$ for $Re_{b}=2400$ for the lower left quadrant of the turbulent square duct flow before transformations.}
\label{fig:b_uv and b_uw}
\end{figure}

\begin{figure}[!ht]
\centering
\subfloat[]{\label{fig:b_uv2400_trans}\includegraphics[width=0.2\linewidth]{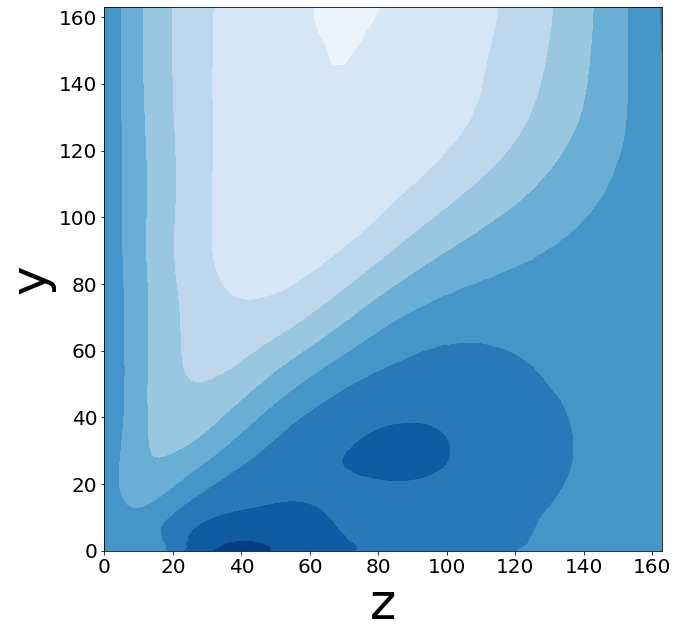}}
\subfloat[]{\label{fig:b_uw2400_trans}\includegraphics[width=0.2\linewidth]{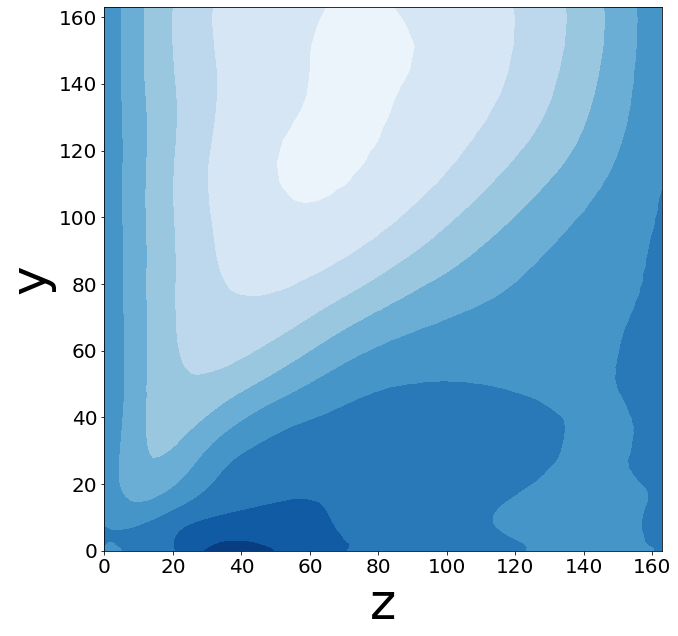}}
\subfloat[]{\label{fig:b_vv2400_trans}\includegraphics[width=0.2\linewidth]{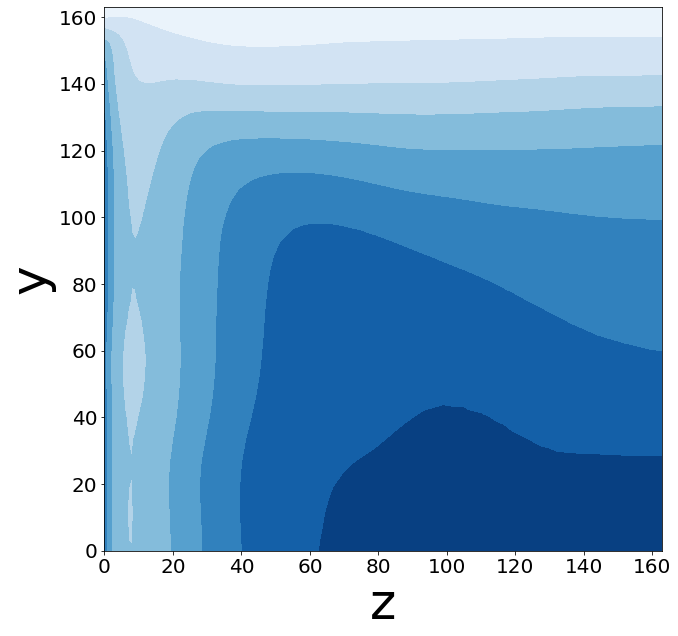}}
\subfloat[]{\label{fig:b_vw2400_trans}\includegraphics[width=0.2\linewidth]{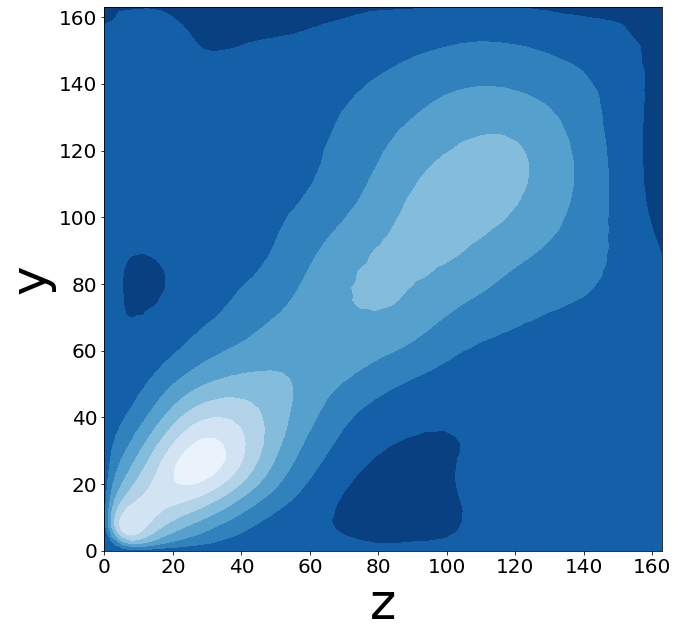}}
\subfloat[]{\label{fig:b_ww2400_trans}\includegraphics[width=0.2\linewidth]{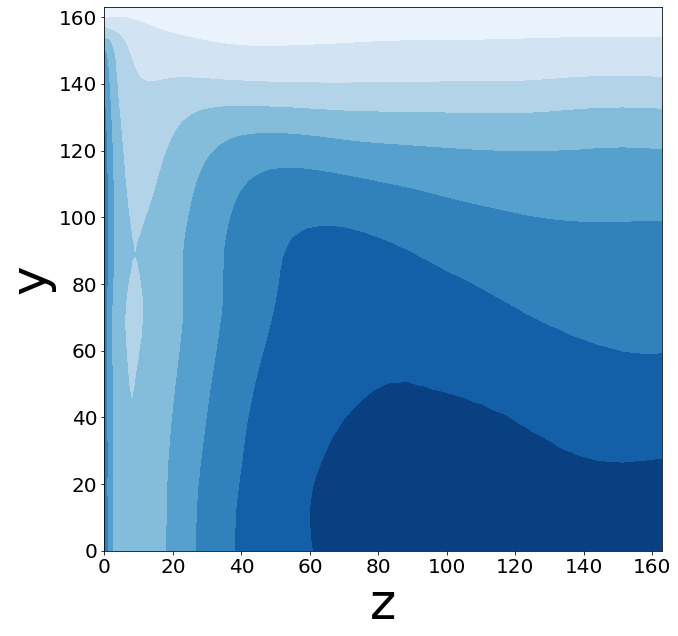}}
\caption{Transformed label data. Qualitative contour plots of (a) $b_{uv}$, (b) $b_{uw}$, (c) $b_{vv}$, (d) $b_{vw}$ and (e) $b_{ww}$ for $Re_{b}=2400$ for the lower left quadrant of the turbulent square duct flow after transformations.}
\label{fig:b_uv and b_uw_transformed}
\end{figure}

Although standard CNN architectures are in general invariant to small distortions, translations, and scaling, they tend to be sensitive to rotations and other deterministic transformations such as mirroring of the data. We aim at constructing an efficient architecture which can be trained as fast as possible using the minimum amount of trainable parameters. To do so, we can apply deterministic transformations to the labels to free the CNN from having to learn transformations such as rotations internally. In this way, we can greatly improve the efficiency of the network and share common ``parent'' layers between different channels of the output $\left(b_{uv},b_{uw},b_{vv},b_{vw},b_{ww}\right)$. Due to the symmetry of the square duct flow, we may expect some of the entries of the Reynolds anisotropic stress tensor to be closely related. For example, Figure~\ref{fig:b_uv2400} and Figure~\ref{fig:b_uw2400} clearly show that $b_{uv}$ and $b_{uw}$ share some qualitative characteristics and spatial features that could be encapsulated in unique filters if it was not for their orientation. The same observation applies to $b_{vv}$ and $b_{ww}$ (see Figure~\ref{fig:b_vv2400} and Figure~\ref{fig:b_ww2400}). We could potentially overcome this issue by tying the weights of groups of filters to several rotated versions of the canonical filter in the group, as suggested by~\cite{GonzalezVT16}. However, in this case, embedding deterministic transformations into the CNN model to ease the algorithm training process and using standard filters seems to be a rather more straightforward answer. All the suggested transformations are summarized in Table~\ref{tab: transformation}. Figure~\ref{fig:b_uv and b_uw} shows qualitative plots of the original labels for $Re_{b}=2400$ and Figure~\ref{fig:b_uv and b_uw_transformed} displays the transformed versions for the same bulk Reynolds number. The practicality of the transformations will become more apparent once we understand the MTLCNN architecture and how MTL is implemented.

\begin{table}[!ht]
\caption{Summary of embedded transformations applied to the training labels for turbulent square duct flow.}
\centering

\begin{tabular}{ll}
\toprule
Parameter & Transformation                                  \\\midrule
$b_{uv}$     & Unchanged                                       \\
$b_{uw}$    & Mirrored about $z$-axis + $270^{\circ}$ clockwise rotation \\
$b_{vv}$     & $90^{\circ}$ clockwise rotation                \\
$b_{vw}$    & Unchanged                                       \\
$b_{ww}$    & Mirrored about $z$-axis \\
\bottomrule   
\label{tab: transformation}
\end{tabular}
\end{table}

\subsection{Main Architecture}
\label{subsec: Main Architecture}

\begin{figure}[htb!]
\centering
\includegraphics[width=\linewidth]{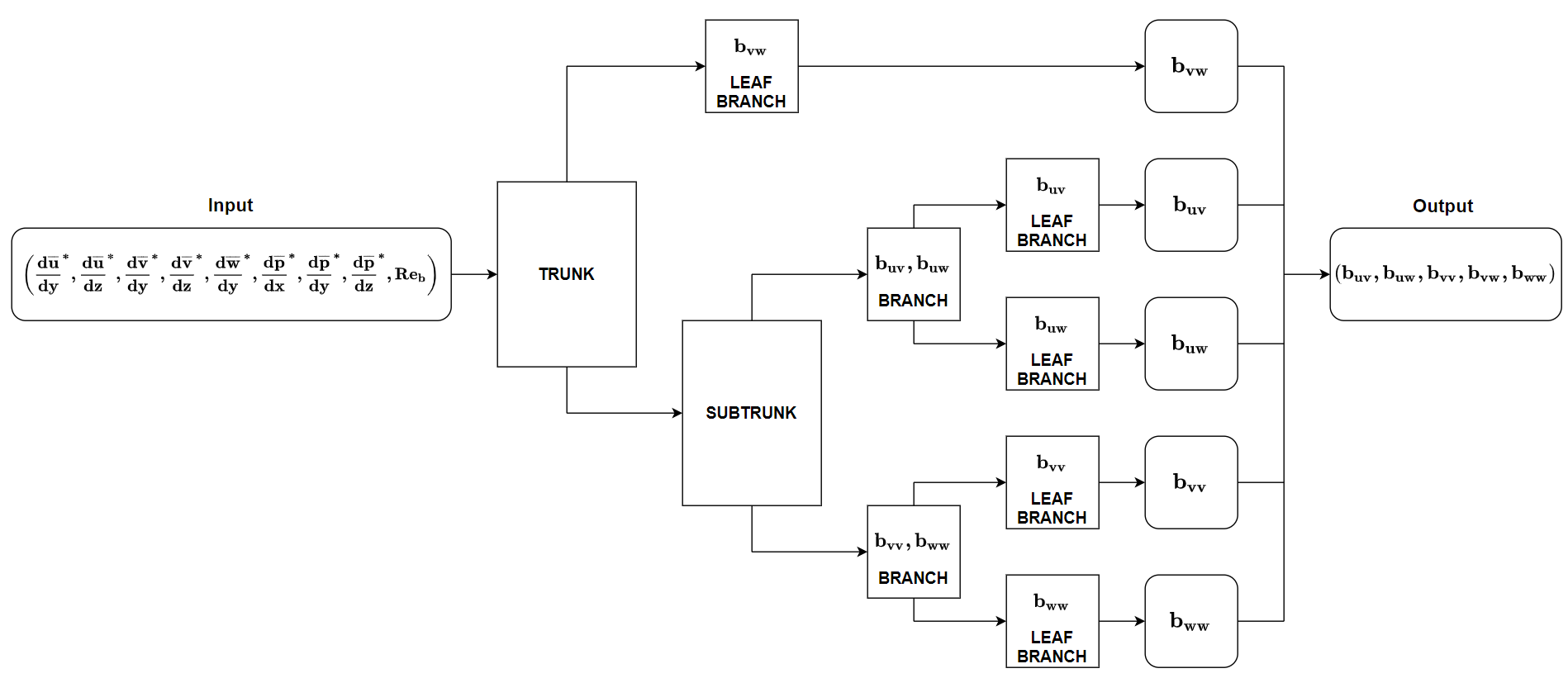}
\caption{Summary of MTLCNN model architecture for turbulent square duct flow.}
\label{fig:summary_of_architecture}
\end{figure}

Figure~\ref{fig:summary_of_architecture} shows a summary of the MTLCNN model which will help better picture the description here presented. Given the assumption that all channels of the model output depend on all channels of the model input, we can intuitively start the model architecture by defining a common set of ``parent'' convolutional layers that share information with all the subsequent layers downstream of the model. We may refer to this first section of the architecture as the ``trunk'' of the MTLCNN model. From Figure~\ref{fig:b_uv and b_uw_transformed}, we can see that different components of the stress tensor have substantially different qualitative characteristics and spatial features. To address this we will ramify the model, which will allow the CNN to accurately predict each component of the anisotropic Reynolds stress tensor. Ramifying consists in dividing the model into different branches or offshoots through which information can flow. This way, different ramifications will learn their own specialized filters for the specific component of the stress tensor that they are trying to capture while still sharing common information from the trunk of the network. 

As suggested by the contour plot in Figure~\ref{fig:b_vw2400_trans}, the $b_{vw}$ component presents the most distinct features. Therefore, the first ramification will aim at predicting $b_{vw}$. The ``$b_{vw}$ leaf branch'' takes as input the activations produced by the last layer of the trunk. Then, it applies several convolutional operations and it computes a trainable weighted sum of the activations of the last convolutional layer in the leaf branch. Next, there is the ``subtrunk''. This part of the network is connected to the trunk but it does not share weights or intervene in the prediction of $b_{vw}$. The subtrunk serves as a bridge between the ``$b_{uv},b_{uw}$ branch'' and the ``$b_{vv},b_{ww}$ branch''. As previously considered in Section~\ref{subsec: Label Transformations}, $b_{uv}$ and $b_{uw}$ share common features as well as $b_{vv}$ and $b_{ww}$. We can think of two groups, the first being $b_{uv}$ and $b_{uw}$ and the second, $b_{vv}$ and $b_{ww}$. At the same time, both groups seem to share lower level features between them. This suggests that all four components $b_{uv}, b_{uw}, b_{vv}$ and $b_{ww}$, can share common convolutional layer filters in the subtrunk and be later ramified into the two groups. Finally, the ``$b_{uv},b_{uw}$ branch'' bifurcates into the ``$b_{uv}$ leaf branch'' and the ``$b_{uw}$ leaf branch''. The same occurs to the ``$b_{vv},b_{ww}$ branch'' which divides into the ``$b_{vv}$ leaf branch'' and the ``$b_{ww}$ leaf branch''. The leaf branches apply extra convolutional operations and compute weighted sum operations based on their final layer outputs. Lastly, all the outputs of the five leaf branches are concatenated into the final prediction, a tensor of dimensions $5\times164\times164$, which is compared to the label data. In total this architecture consists of 56,660 trainable parameters. Table~\ref{tab:model_summary} reviews all the layers in the model, describes how they are connected between them, specifies the number and size of the convolutional filters that they use and includes other relevant information.

Using common ``parent'' layers upstream of the model allows their filters to learn basic relationships which are shared by all components of the normalized Reynolds anisotropic stress tensor. Later, the ramifications help filters downstream of the model to learn relationships which are specific to certain components of the prediction. The weights of layers downstream of the model are only updated based on part of the model output. In essence, the overall loss of the MTLCNN model can be subdivided into five losses, one for each component of the model prediction. The weights of the leaf branches are only updated based on a single loss, whereas the weights of the trunk are affected by all losses. Note that we use zero padding throughout the MTLCNN model, which means that the input, activations and output always have dimensions $\,c\times164\times164$, where $c$ is the channel dimension. We would like to highlight that addressing this problem without using MTL, ramifications, and label transformations becomes especially challenging because the number of required trainable parameters skyrockets and training the algorithm becomes extremely computationally expensive.

\begin{sidewaystable}[]
\caption[Summary of CNN model architecture for turbulent square duct flow.]{Summary of CNN model architecture for turbulent square duct flow. On the leftmost column the layer name is specified. The table includes information regarding the connectivity between layers, the number of filters and kernel size used by each convolutional layer, whether or not batch normalization is applied to the output of a given layer and the activation function used.}
\resizebox{\linewidth}{!}{

\begin{tabular}{lllcccc}
\toprule
LAYER                                     & \multicolumn{1}{l}{TAKES INPUT FROM}                               & \multicolumn{1}{l}{GIVES OUTPUT TO}                            & \multicolumn{1}{l}{N$^{\circ}$ FILTERS} & \multicolumn{1}{l}{KERNEL SIZE} & \multicolumn{1}{l}{BATCH NORMALIZATION} & \multicolumn{1}{l}{ACTIVATION} \\\midrule
Trunk L1                             & Model input                                                         & Trunk L2                                                   & 9                                     & 3                                            & Yes                                      & ELU                                      \\
Trunk L2                             & Trunk L1                                                       & Trunk L3                                                   & 9                                     & 3                                            & Yes                                      & ELU                                      \\
Trunk L3                             & Trunk L2                                                       & Trunk L4                                                   & 9                                     & 3                                            & Yes                                      & ELU                                      \\
Trunk L4                             & Trunk L3                                                       & Trunk L5 & 9                                     & 3                                            & Yes                                      & ELU                                      \\Trunk L5                             & Trunk L4                                                       & $b_{vw}$ Leaf Branch L1 and Subtrunk L1 & 9                                     & 3                                            & Yes                                      & ELU                                      \\ \midrule
$b_{vw}$ Leaf Branch L1                      & Trunk L5                                                       & $b_{vw}$ Leaf Branch L2                                            & 10                                     & 3                                            & Yes                                      & ELU                                      \\ 
$b_{vw}$ Leaf Branch L2                      & $b_{vw}$ Leaf Branch L1                                                & $b_{vw}$ Leaf Branch L3                                            & 10                                     & 3                                            & Yes                                      & ELU                                      \\ 
$b_{vw}$ Leaf Branch L3                      & $b_{vw}$ Leaf Branch L2                                                & $b_{vw}$ Leaf Branch L4                                            & 10                                     & 3                                            & Yes                                      & ELU                                      \\ 
$b_{vw}$ Leaf Branch L4                      & $b_{vw}$ Leaf Branch L3                                                & $b_{vw}$ Leaf Branch L5                                            & 10                                     & 3                                            & Yes                                      & ELU                                      \\ 
$b_{vw}$ Leaf Branch L5                      & $b_{vw}$ Leaf Branch L4                                                & $b_{vw}$ Leaf Branch L6                                       & 10                                     & 3                                            & Yes                                       & ELU                                      \\ 
$b_{vw}$ Leaf Branch L6                      & $b_{vw}$ Leaf Branch L5                                                & $b_{vw}$ Leaf Branch L7                                       & 15                                     & 3                                            & Yes                                       & ELU                                      \\ 
$b_{vw}$ Leaf Branch L7                      & $b_{vw}$ Leaf Branch L6                                                & $b_{vw}$ Leaf Branch Weighted Sum                                       & 35                                     & 3                                            & No                                       & ELU                                      \\ 
$b_{vw}$ Leaf Branch Weighted Sum                 & $b_{vw}$ Leaf Branch L7                                                & Concatenate Output                                              & -                                      & -                                            & No                                       & None                                     \\ \midrule
Subtrunk L1 & Trunk L5                                                       & Subtrunk L2                       & 10                                     & 3                                            & Yes                                      & ELU                                      \\
Subtrunk L2 & Subtrunk L1                           & $b_{uv}$, $b_{uw}$ Branch L1 and $b_{vv}$, $b_{ww}$ Branch L1     & 10                                     & 3                                            & Yes                                      & ELU                                      \\ \midrule
$b_{uv}$, $b_{uw}$ Branch L1               & Subtrunk L2                           & $b_{uv}$, $b_{uw}$ Branch L2                                     & 10                                     & 3                                            & Yes                                      & ELU                                      \\ 
$b_{uv}$, $b_{uw}$ Branch L2               & $b_{uv}$, $b_{uw}$ Branch L1                                         & $b_{uv}$ Leaf Branch L1 and $b_{uw}$ Leaf Branch L1                   & 10                                     & 3                                            & Yes                                      & ELU                                      \\ \midrule
$b_{uv}$ Leaf Branch L1                      & $b_{uv}$, $b_{uw}$ Branch L2                                         & $b_{uv}$ Leaf Branch L2                                            & 10                                     & 3                                            & Yes                                      & ELU                                      \\ 
$b_{uv}$ Leaf Branch L2                      & $b_{uv}$ Leaf Branch L1                                                & $b_{uv}$ Leaf Branch L3                                            & 20                                     & 3                                            & Yes                                      & ELU                                      \\ 
$b_{uv}$ Leaf Branch L3                      & $b_{uv}$ Leaf Branch L2                                                & $b_{uv}$ Leaf Branch Weighted Sum                                       & 35                                     & 3                                            & No                                       & ELU                                      \\ 
$b_{uv}$ Leaf Branch Weighted Sum                 & $b_{uv}$ Leaf Branch L3                                                & Concatenate Output                                              & -                                      & -                                            & No                                       & None                                     \\ \midrule
$b_{uw}$ Leaf Branch L1                      & $b_{uv}$, $b_{uw}$ Branch L2                                         & $b_{uw}$ Leaf Branch L2                                            & 10                                     & 3                                            & Yes                                      & ELU                                      \\ 
$b_{uw}$ Leaf Branch L2                      & $b_{uw}$ Leaf Branch L1                                                & $b_{uw}$ Leaf Branch L3                                            & 20                                     & 3                                            & Yes                                      & ELU                                      \\ 
$b_{uw}$ Leaf Branch L3                      & $b_{uw}$ Leaf Branch L2                                                & $b_{uw}$ Leaf Branch Weighted Sum                                               & 35                                     & 3                                            & No                                       & ELU                                      \\ 
$b_{uw}$ Leaf Branch Weighted Sum                 & $b_{uw}$ Leaf Branch L3                                                & Concatenate Output                                              & -                                      & -                                            & No                                       & None                                     \\ \midrule
$b_{vv}$, $b_{ww}$ Branch L1               & Subtrunk L2                           & $b_{vv}$, $b_{ww}$ Branch L2                                     & 10                                     & 3                                            & Yes                                      & ELU                                      \\ 
$b_{vv}$, $b_{ww}$ Branch L2               & $b_{vv}$, $b_{ww}$ Branch L1                                         & $b_{vv}$ Leaf Branch L1 and $b_{ww}$ Leaf Branch L1                   & 10                                     & 3                                            & Yes                                      & ELU                                      \\ \midrule
$b_{vv}$ Leaf Branch L1                      & $b_{vv}$, $b_{ww}$ Branch L2                                         & $b_{vv}$ Leaf Branch L2                                            & 10                                     & 3                                            & Yes                                      & ELU                                      \\ 
$b_{vv}$ Leaf Branch L2                      & $b_{vv}$ Leaf Branch L1                                                & $b_{vv}$ Leaf Branch L3                                            & 20                                     & 3                                            & Yes                                      & ELU                                      \\ 
$b_{vv}$ Leaf Branch L3                      & $b_{vv}$ Leaf Branch L2                                                & $b_{vv}$ Leaf Branch Weighted Sum                                       & 35                                     & 3                                            & No                                       & ELU                                      \\ 
$b_{vv}$ Leaf Branch Weighted Sum                        & $b_{vv}$ Leaf Branch L3                                                & Concatenate Output                                              & -                                      & -                                            & No                                       & None                                     \\ \midrule
$b_{ww}$ Leaf Branch L1                      & $b_{vv}$, $b_{ww}$ Branch L1                                         & $b_{ww}$ Leaf Branch L2                                            & 10                                     & 3                                            & Yes                                      & ELU                                      \\ 
$b_{ww}$ Leaf Branch L2                      & $b_{ww}$ Leaf Branch L1                                                & $b_{ww}$ Leaf Branch L3                                            & 20                                     & 3                                            & Yes                                      & ELU                                      \\ 
$b_{ww}$ Leaf Branch L3                      & $b_{ww}$ Leaf Branch L2                                                & $b_{ww}$ Leaf Branch Weighted Sum                                       & 35                                     & 3                                            & No                                       & ELU                                      \\ 
$b_{ww}$ Leaf Branch Weighted Sum                 & $b_{ww}$ Leaf Branch L3                                                & Concatenate Output                                              & -                                      & -                                            & No                                       & None                                      \\ \midrule
Concatenate Output                        & $b_{uv}$, $b_{uw}$, $b_{vv}$, $b_{vw}$ and $b_{ww}$ Leaf Branches Weighted Sum & Model Output                                                    & -                                      & -                                            & No                                       & None                                     \\ \bottomrule
\end{tabular}
}
\label{tab:model_summary}
\end{sidewaystable}

\subsection{Filtering and Boosting}
\label{subsec: Filtering and Boosting}

Once the model has already been trained using the architecture described in Section~\ref{subsec: Main Architecture}, we further improve the prediction by adding a series of operations to the model output after freezing the weights of the MTLCNN. Convolution in neural networks is an operation that to some degree resembles a scanner, which can lead to border effects and meaningless artifacts in the solution. When the kernel slides over the borders of the input we use zero padding to preserve the original input size, but this means that  there is effectively ``less information'' for the filter to capture close to the borders. We find that the predictions made by the model may sometimes present meaningless sudden variations close to the borders of the output. We try to address this problem by using a special loss function during training that will be discussed in Section~\ref{sec:Results}. However, we can further improve the model performance by smoothing the prediction using a uniform filter; a square filter of size 20 was found to give the best results. The filter is applied to each component prediction independently (to each $164\times164$ image separately). Note that the input array is extended by replicating the last pixel when the smoothing filter overlaps a border. Once we have filtered the image, we obtain a more realistic and smooth prediction. Although the prediction is better from a qualitative perspective (it does not present sudden meaningless changes), this operation negatively affects the accuracy of the model. We regain performance by applying the method of least squares for each entry of the $5\times164\times164$ prediction across all examples in the training data, that is, based on the data for the bulk Reynolds numbers available for training. 
The least square loss is
\begin{equation}
    \ell(w_{b})=\frac{1}{n} \sum_{i=1}^{n}\left(\gamma_{i}-w_{b}\beta_{i} \right)^{2},
\end{equation}
where $n$ is the number of training examples ($n=15$ see Section~\ref{sec:Results}), $\gamma_{i}$ is the corresponding value of the entry in the training label data, $\beta_{i}$ is the original model prediction, and $w_{b}$ is the weight we are trying to find. We find that overall applying the operations described in this section helps boost the model performance from a qualitative perspective, since it gives more realistic predictions in which the effect of artifacts is attenuated. However, note that filtering and boosting is not key to our method, but simply convenient.

\section{Results}
\label{sec:Results}

Next, we discuss the training procedure for the MTLCNN model. Before training, we use the Kaiming He uniform distribution to initialize the weights and biases of the convolutional layers in the network (see~\cite{he2015delving}). On the other hand, the weights of the weighted sum operations are initialized at $\frac{1}{35}$. 
Using a constant learning rate was found to produce large oscillations in the training loss and hinders convergence. 
To circumvent this issue, the learning was adapted based on the epoch according to 
\begin{equation}
    \eta=10^{-3}\times\left(0.9975^{\text{epoch}}\right)+2\times10^{-3}\times\left(0.2^{\text{epoch}}\right)\,,
    \label{eq:learning_rate}
\end{equation}
where $\eta$ is the learning rate. This scheme decreases the learning rate as the number of epochs increases. When the training process starts the weights are initially random and taking bigger steps helps speed up convergence. As the number of epochs increases and the training loss decreases and starts converging towards a minimum, we benefit from the reduction in learning rate. A batch size of $1$ is used during training because the model is updated based on the data for one bulk Reynolds number at a time. 
Using the standard MSE to train this network yielded results with excessive sudden changes between nearby entries of the output. 
To address this issue, a the loss function was modified with a regularization term,
\begin{equation}
    L^{\prime}(\boldsymbol{\theta})=\text{MSE}\left(\mathbf{y}^{\text{true}},\mathbf{y}(\boldsymbol{\theta})\right)+\vartheta\sum_{k=1}^{5}\left(\sum_{i=1}^{163}\lvert y_{k(i+1)j}(\boldsymbol{\theta})-y_{kij}(\boldsymbol{\theta})\rvert +\sum_{j=1}^{163}\lvert y_{ki(j+1)}(\boldsymbol{\theta})-y_{kij}(\boldsymbol{\theta})\rvert\right)\,,
    \label{eq:modified_mse}
\end{equation}
where $\boldsymbol{\theta}$ represents the model parameters, $\mathbf{y}^{\text{true}}$ the true normalized anisotropic Reynolds stress tensor, $\mathbf{y}$ the prediction, and $y_{kij}$ is a single entry of the output tensor, $k$ being the channel dimension, and $i$ and $j$ the subscripts representing the height and width dimensions of the output ($y$ and $z$ dimensions on the cross-sectional area of the duct). $\vartheta$ is a hyperparameter that regulates the contribution of the second term to the loss function. A value of $\vartheta=10^{-7}$ was found to give the best trade-off between accuracy and smoothness of the solution. 
In this way, the loss function $L^{\prime}(\boldsymbol{\theta})$ accounts for the MSE between the label and predicted anisotropic Reynolds stress tensor while trying to find a solution without large changes between adjacent entries of the prediction. 



As discussed in Section~\ref{subsec:Turbulent Duct flow}, the dataset by~\cite{mcconkey2021curated} has data for turbulent duct flow ranging between $Re_{b}=1100$ and $Re_{b}=3500$. In total, there are 16 bulk Reynolds numbers. We will define 16 training cases in which we train the model with 15 of the bulk Reynolds numbers and test its performance on the data for the remaining bulk Reynolds number, as shown in Table~\ref{tab:training-prediction cases duct flow}.

\begin{table}[htb!]
\centering
\caption{The 16 training-prediction cases for turbulent square duct flow.}
\label{tab:training-prediction cases duct flow}
\resizebox{\columnwidth}{!}{
\begin{tabular}{@{}ccc@{}}
\toprule
Case & Training set            & Test set \\ \midrule
1    & Re$_{b}={[}1100,1150,1250,1300,1350,1400,1500,1600,1800,2000,2205,2400,2600,2900,3200{]}$  & Re$_{b}=3500$  \\
2    & Re$_{b}={[}1100,1150,1250,1300,1350,1400,1500,1600,1800,2000,2205,2400,2600,2900,3500{]}$  & Re$_{b}=3200$  \\
3    & Re$_{b}={[}1100,1150,1250,1300,1350,1400,1500,1600,1800,2000,2205,2400,2600,3200,3500{]}$  & Re$_{b}=2900$  \\
4    & Re$_{b}={[}1100,1150,1250,1300,1350,1400,1500,1600,1800,2000,2205,2400,2900,3200,3500{]}$ & Re$_{b}=2600$ \\
5    & Re$_{b}={[}1100,1150,1250,1300,1350,1400,1500,1600,1800,2000,2205,2600,2900,3200,3500{]}$ & Re$_{b}=2400$ \\
6    & Re$_{b}={[}1100,1150,1250,1300,1350,1400,1500,1600,1800,2000,2400,2600,2900,3200,3500{]}$ & Re$_{b}=2205$ \\
7    & Re$_{b}={[}1100,1150,1250,1300,1350,1400,1500,1600,1800,2205,2400,2600,2900,3200,3500{]}$ & Re$_{b}=2000$ \\
8    & Re$_{b}={[}1100,1150,1250,1300,1350,1400,1500,1600,2000,2205,2400,2600,2900,3200,3500{]}$ & Re$_{b}=1800$ \\
9    & Re$_{b}={[}1100,1150,1250,1300,1350,1400,1500,1800,2000,2205,2400,2600,2900,3200,3500{]}$ & Re$_{b}=1600$ \\
10    & Re$_{b}={[}1100,1150,1250,1300,1350,1400,1600,1800,2000,2205,2400,2600,2900,3200,3500{]}$ & Re$_{b}=1500$ \\
11   & Re$_{b}={[}1100,1150,1250,1300,1350,1500, 1600,1800,2000,2205,2400,2600,2900,3200,3500{]}$ & Re$_{b}=1400$ \\
12   & Re$_{b}={[}1100,1150,1250,1300,1400,1500, 1600,1800,2000,2205,2400,2600,2900,3200,3500{]}$ & Re$_{b}=1350$ \\
13   & Re$_{b}={[}1100,1150,1250,1350,1400,1500, 1600,1800,2000,2205,2400,2600,2900,3200,3500{]}$ & Re$_{b}=1300$ \\
14   & Re$_{b}={[}1100,1150,1300,1350,1400,1500, 1600,1800,2000,2205,2400,2600,2900,3200,3500{]}$ & Re$_{b}=1250$ \\
15   & Re$_{b}={[}1100,1250,1300,1350,1400,1500, 1600,1800,2000,2205,2400,2600,2900,3200,3500{]}$ & Re$_{b}=1150$ \\
16   & Re$_{b}={[}1150,1250,1300,1350,1400,1500, 1600,1800,2000,2205,2400,2600,2900,3200,3500{]}$ & Re$_{b}=1100$ \\
\bottomrule
\end{tabular}}
\end{table}

\begin{figure}[htbp!]
\centering
\subfloat[]{\label{fig:vw_3500}\includegraphics[width=0.25\linewidth]{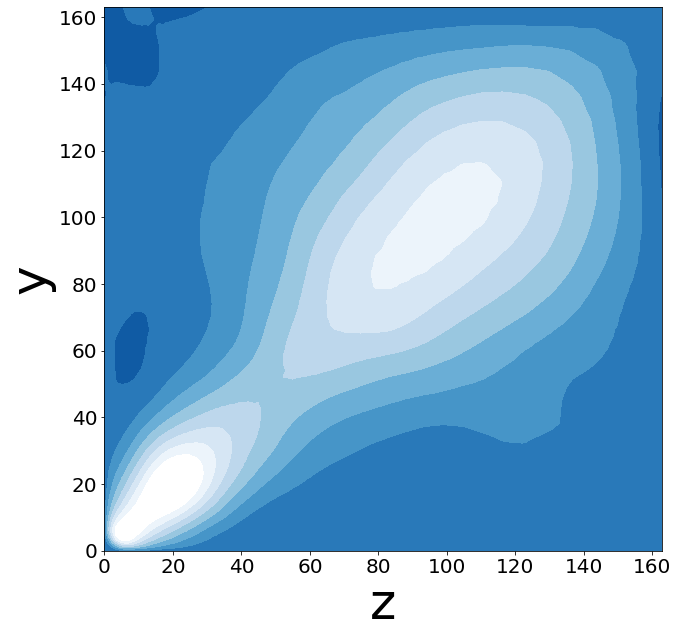}}
\subfloat[]{\label{fig:vw_2205}\includegraphics[width=0.25\linewidth]{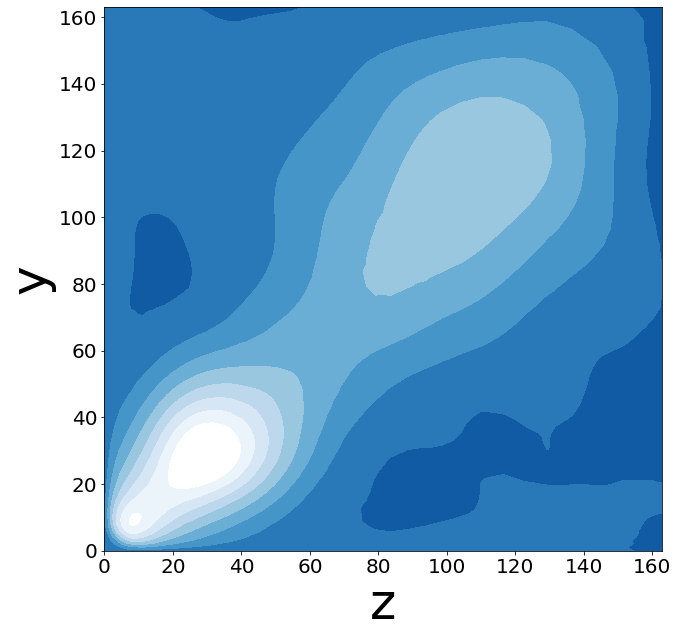}}
\subfloat[]{\label{fig:vw_1400}\includegraphics[width=0.25\linewidth]{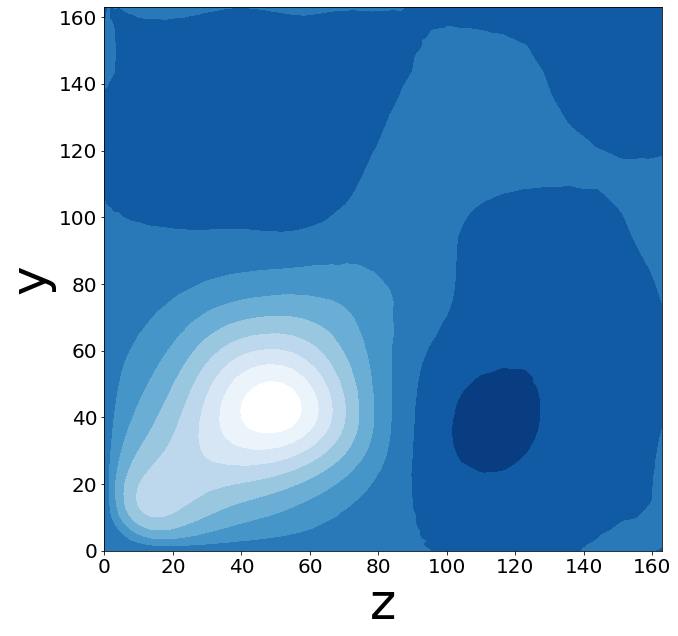}}
\subfloat[]{\label{fig:vw_1100}\includegraphics[width=0.25\linewidth]{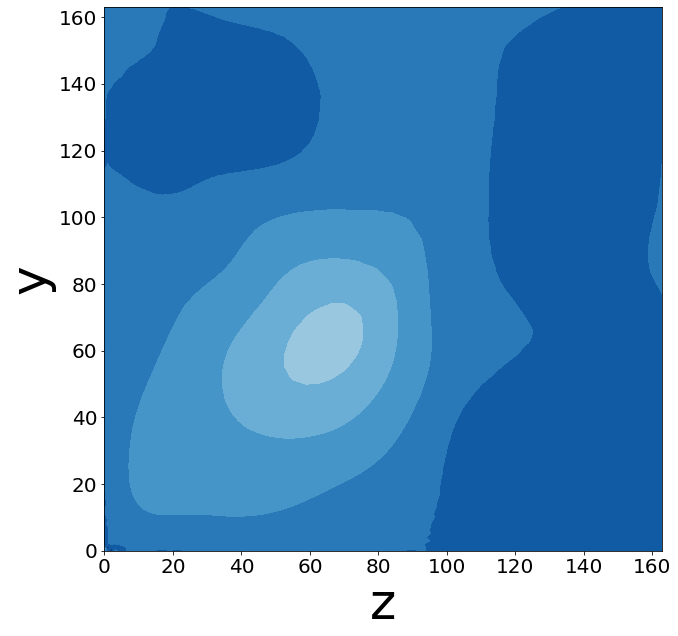}}
\caption{Qualitative contour plots of labels for $b_{vw}$ at (a) $Re_{b}=3500$, (b) $Re_{b}=2205$, (c) $Re_{b}=1400$ and (e) $Re_{b}=1100$.}
\label{fig:vw_for_different_Reynolds}
\end{figure}

The standard training approach would consist in feeding all the 16 examples to the CNN model each epoch, so that the model is updated 16 times per epoch and tries to learn the normalized anisotropic Reynolds stress tensor for all bulk Reynolds numbers from $Re_{b}=1100$ to $Re_{b}=3500$ from the start. Indeed, this training procedure does work and gives good results. Nevertheless, it is inefficient from a computational perspective. As we progress into developing turbulence models that can be employed to predict more complex flows we can expect the computational cost of such models to increase. It is of great importance that we propose less expensive training methods for the neural network models. We can speed up the convergence of the algorithm by designing a training curricula. Note that at $Re_{b}=1100$ turbulent duct flow is at a marginal state, since it is the lowest bulk Reynolds number for which self-sustaining turbulence can be achieved, according to simulations by~\cite{marginal}. Hence, we expect the CNN model to find more complications for predictions at lower bulk Reynolds numbers, because the flow is more sensitive to variations in $Re_{b}$. We can corroborate this fact by inspecting the DNS labels for different bulk Reynolds numbers. Figure~\ref{fig:vw_for_different_Reynolds} displays the $b_{vw}$ component plot for different bulk Reynolds numbers, with the highest bulk Reynolds number $Re_{b}=3500$ starting on the left and the lowest $Re_{b}=1100$ on the right. There is a greater difference in bulk Reynolds number between Figure~\ref{fig:vw_3500} ($Re_{b}=3500$) and Figure~\ref{fig:vw_2205} ($Re_{b}=2205$) than there is between Figure~\ref{fig:vw_2205} ($Re_{b}=2205$) and Figure~\ref{fig:vw_1100} ($Re_{b}=1100$), but we can see that the $b_{vw}$ component changes more drastically as we approach low bulk Reynolds numbers, as expected. This trend is consistent for all other components of the normalized anisotropic Reynolds stress tensor too. We can take this fact into account when applying curriculum learning. Considering the physics-based intuition of the flow, we may start by training the model with data for high bulk Reynolds numbers first and gradually include examples closer to the marginal state of the flow as training progresses. Therefore, we may divide the training process into three steps: first we train the model for the 5 highest $Re_{b}$ cases for 50 epochs, then we move on to train on the 10 highest $Re_{b}$ for 200 epochs and finally we train the model on all the cases for the remaining number of epochs. In each step we update the model based on the highest $Re_{b}$ first.

\begin{figure}[!ht]
        \centering
        \begin{subfigure}[b]{0.48\textwidth}
            \centering
            \includegraphics[width=\textwidth]{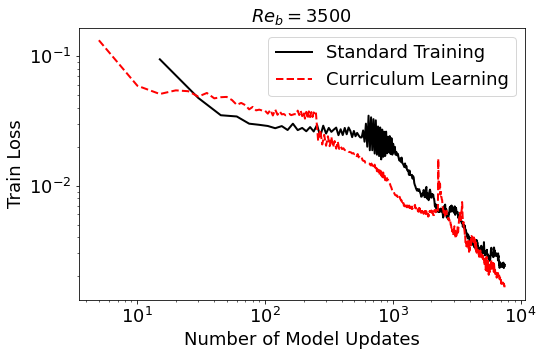}
            \caption[]%
            {{}}    
            \label{Train_cl_3500}
        \end{subfigure}
        \begin{subfigure}[b]{0.48\textwidth}  
            \centering 
            \includegraphics[width=\textwidth]{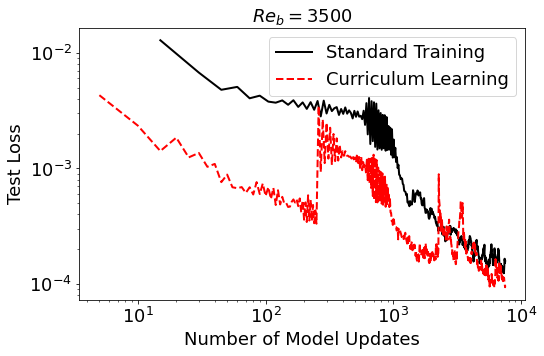}
            \caption[]%
            {{}}    
            \label{Test_cl_3500}
        \end{subfigure}
        \vskip\baselineskip
        \begin{subfigure}[b]{0.48\textwidth}   
            \centering 
            \includegraphics[width=\textwidth]{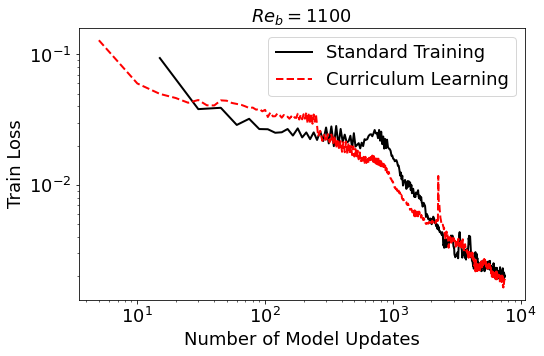}
            \caption[]%
            {{}}    
            \label{Train_cl_1100}
        \end{subfigure}
        \begin{subfigure}[b]{0.48\textwidth}   
            \centering 
            \includegraphics[width=\textwidth]{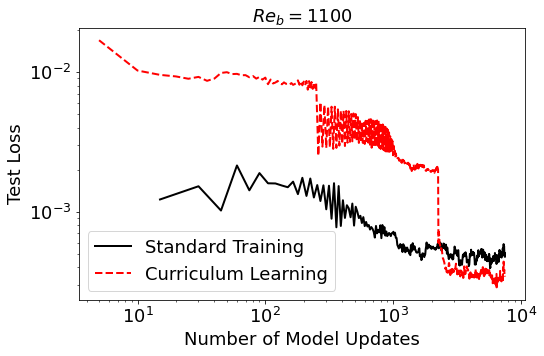}
            \caption[]%
            {{}}    
            \label{Test_cl_1100}
        \end{subfigure}
        \caption[Loss function against number of model updates when using standard training vs curriculum learning.]
        {Loss function against number of model updates when using standard training vs curriculum learning. (a) Train loss for $Re_{b}$=3500. (b) Test loss for $Re_{b}$=3500. (c) Train loss for $Re_{b}$=1100. (d) Test loss for $Re_{b}$=1100.} 
        \label{fig:curriculum_learning_comparisons}
\end{figure}

Figure~\ref{fig:curriculum_learning_comparisons} shows a comparison between the evolution of the loss for standard training and curriculum learning for both $Re_{b}=3500$ and $Re_{b}=1100$, the highest and lowest bulk Reynolds numbers in the dataset. The loss is plotted against the number of model updates instead of the number of epochs to allow for a fair comparison. The plots for curriculum learning start slightly before the ones for standard training. This is because in the case of curriculum learning the model is updated five times in the first epoch, whereas in standard training it is updated fifteen times, and the loss function is only recorded at the end of each epoch. Note that we use the same random seed when training the algorithm, so that we can compare training using the standard approach and curriculum learning both starting from the same random weight distribution. As we can see from the plots, curriculum learning outperforms standard training. Curriculum learning achieves a better training loss (Figure~\ref{Train_cl_3500} and Figure~\ref{Train_cl_1100}) and it improves generalization (Figure~\ref{Test_cl_3500} and Figure~\ref{Test_cl_1100}), which is in line with previous findings by~\cite{cl_bengio}. Figure~\ref{Test_cl_1100} is particularly interesting. During the first two steps of the training curricula the algorithm is only trained on bulk Reynolds numbers between $Re_{b}=3500$ and $Re_{b}=1500$, and the test loss function using curriculum learning stays above the line for standard training. However, once we reach step three of curriculum learning in which we train the model on all the bulk Reynolds numbers in the train set, the test loss falls abruptly below the loss achieved using standard training. Indeed, in Figure~\ref{Test_cl_1100} we can see how learning based on a smooth version of the loss function first has helped us reach a better minimum for the target loss.

In Table~\ref{curriculum_learning_std_training_table} we can find a summary of the model results (before filtering and boosting) obtained using standard training and curriculum learning. The $R^{2}$ error obtained using curriculum learning is better, although not excessively. This can be attributed to the fact that the neural network architecture already has a really good performance so the marginal room for improvement is small. However, this finding suggests that using curriculum learning may be beneficial to use in future more complex problems where it could possibly be a bigger determining factor for the model performance and reduce computational time.

\begin{figure}[!ht]
\centering
\subfloat[]{\label{fig:before1800uv}\includegraphics[height=0.35\linewidth]{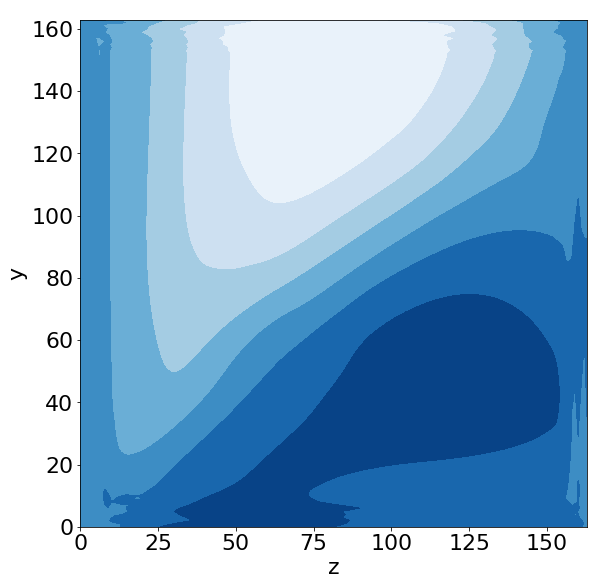}}
\subfloat[]{\label{fig:after1800uv}\includegraphics[height=0.35\linewidth]{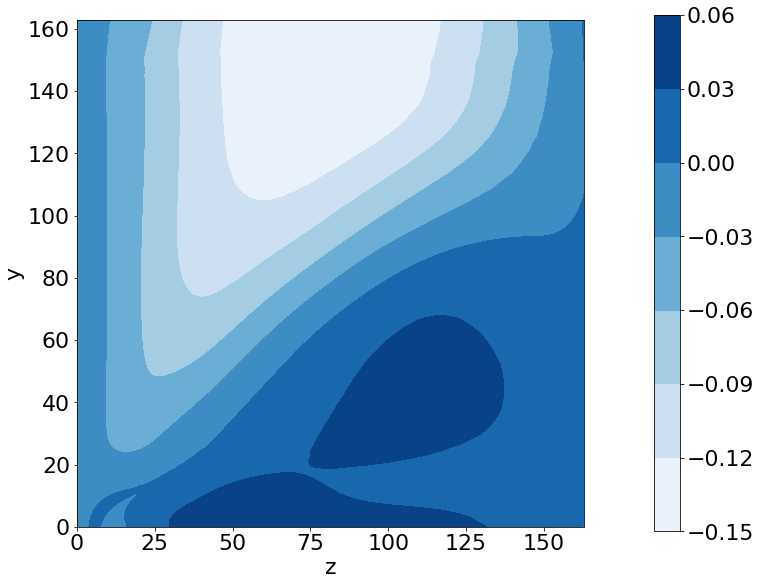}}
\caption{Model prediction of $b_{uv}$ for Case 8 testing on $Re_{b}=1800$ (a) before and (b) after filtering and boosting.}
\label{fig:before and after 1800 uv}
\end{figure}

Lastly, as discussed in Section~\ref{subsec: Filtering and Boosting}, we apply filtering and boosting to improve the performance of the model and get rid of meaningless artifacts that may appear in some of the solutions. Table~\ref{curriculum_learning_std_training_table} displays a comparison between the $R^{2}$ errors obtained before filtering and boosting and after. The filtering and boosting operations are applied to the model results obtained using curriculum learning. From the results we find that filtering and boosting do in most of the cases help with the generalization of the model and improve the performance on the training and testing data. Figure~\ref{fig:before and after 1800 uv} shows an example comparison between the original model output and the final result after filtering and boosting. We can see that filtering and boosting gives a more reliable solution without artifacts.

\begin{figure}[!ht]
\centering
\subfloat[Prediction $b_{uv}$]{\label{fig:before1350uv}\includegraphics[height=0.25\linewidth]{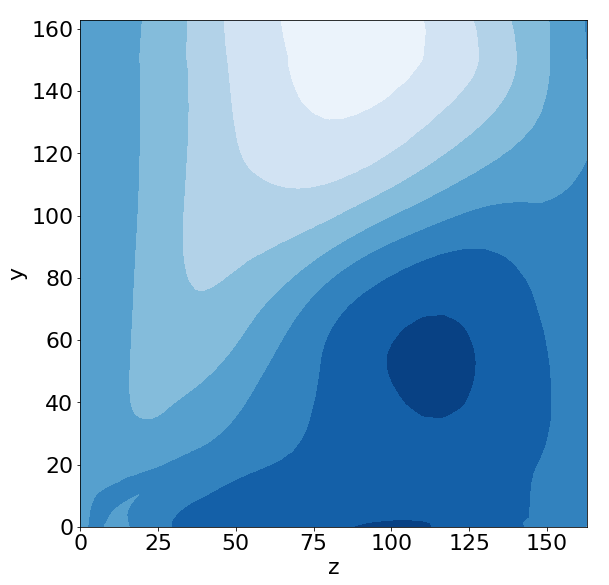}}
\subfloat[DNS $b_{uv}$]{\label{fig:after1350uv}\includegraphics[height=0.25\linewidth]{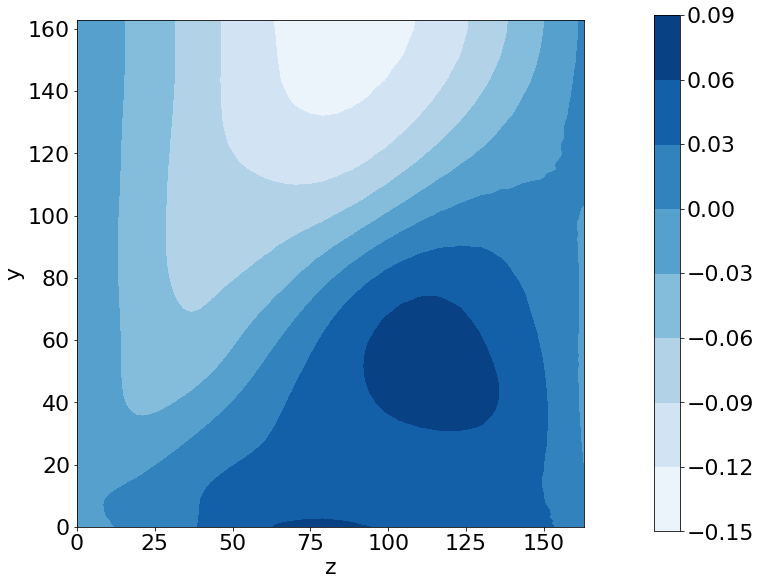}}
\\
\subfloat[Prediction $b_{uw}$]{\label{fig:before1350uw}\includegraphics[height=0.25\linewidth]{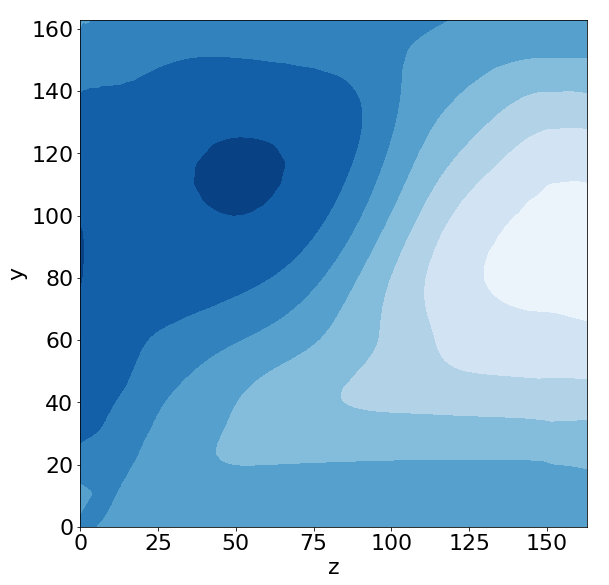}}
\subfloat[DNS $b_{uw}$]{\label{fig:after1350uw}\includegraphics[height=0.25\linewidth]{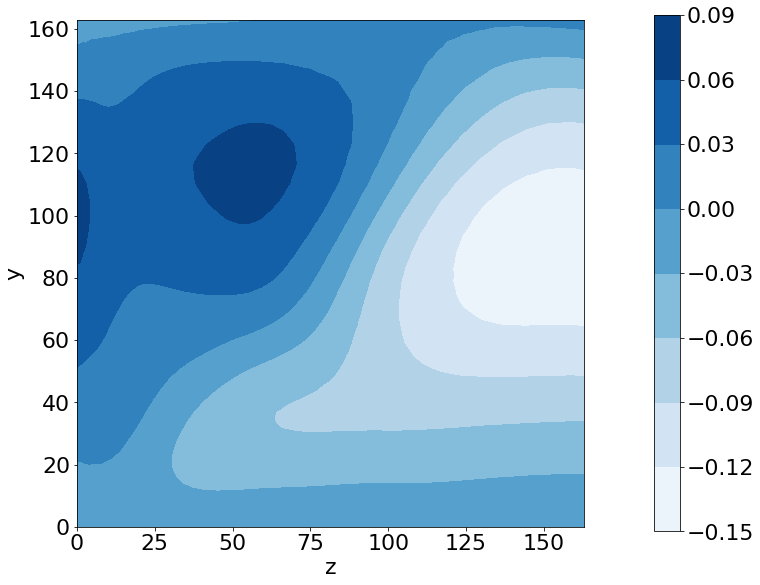}}
\\
\subfloat[Prediction $b_{vv}$]{\label{fig:before1350vv}\includegraphics[height=0.25\linewidth]{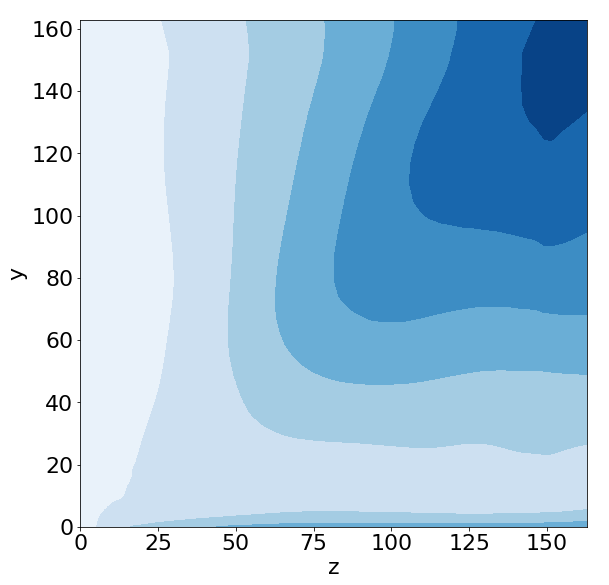}}
\subfloat[DNS $b_{vv}$]{\label{fig:after1350vv}\includegraphics[height=0.25\linewidth]{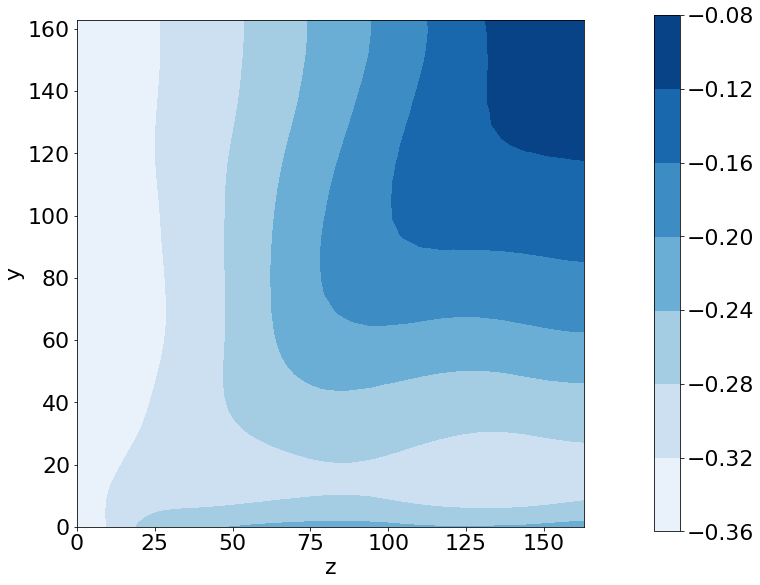}}
\\
\subfloat[Prediction $b_{vw}$]{\label{fig:before1350vw}\includegraphics[height=0.25\linewidth]{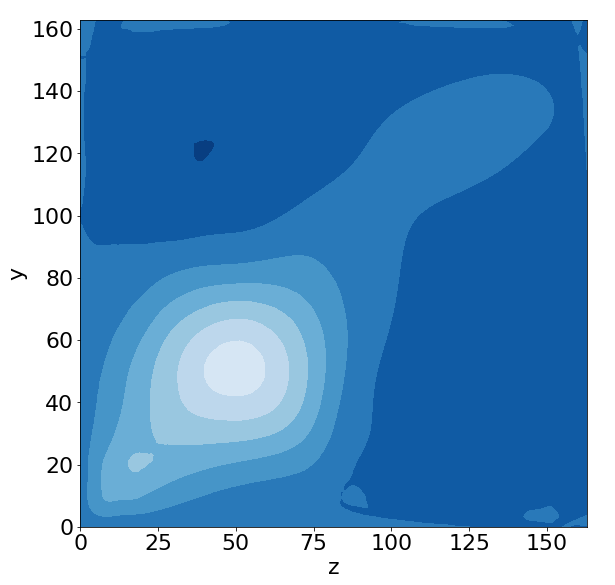}}
\subfloat[DNS $b_{vw}$]{\label{fig:after1350vw}\includegraphics[height=0.25\linewidth]{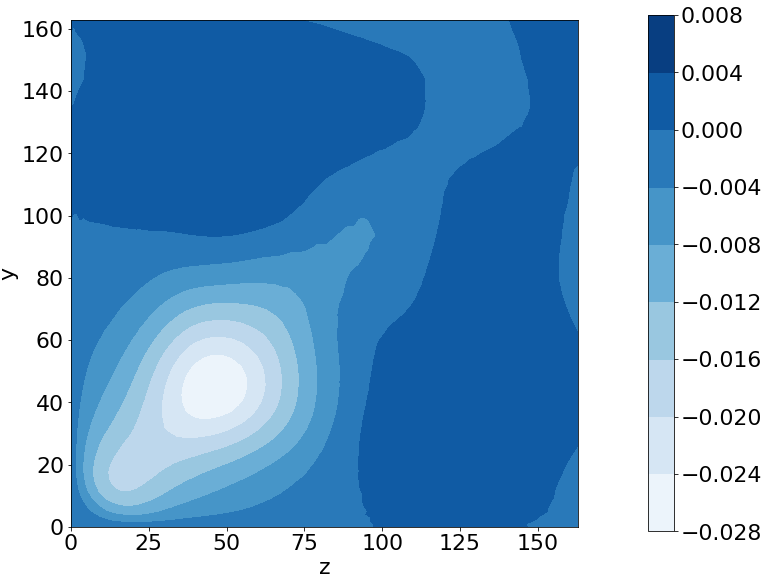}}
\\
\subfloat[Prediction $b_{ww}$]{\label{fig:before1350ww}\includegraphics[height=0.25\linewidth]{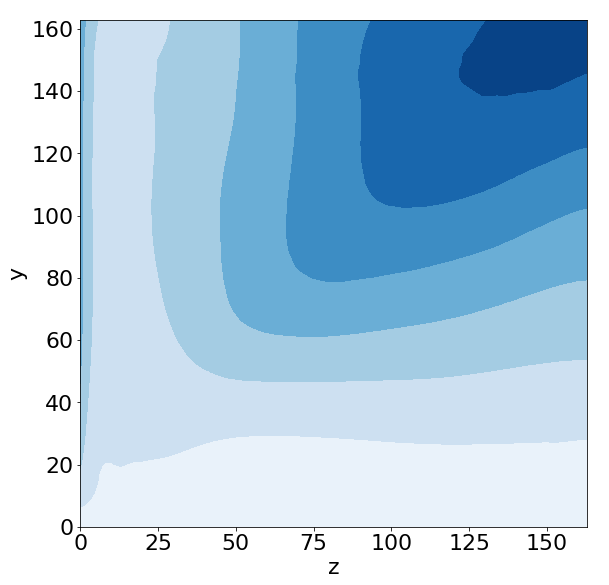}}
\subfloat[DNS $b_{ww}$]{\label{fig:after1350ww}\includegraphics[height=0.25\linewidth]{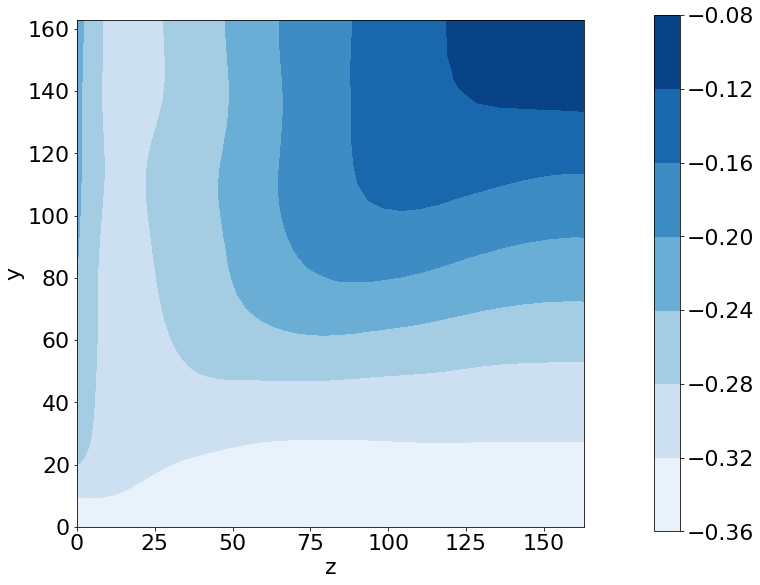}}

\caption{Square duct flow final model predictions and label DNS data for $Re_{b}=1350$ (Case 12).}
\label{fig:prediction1350uv}
\end{figure}

Overall, we find that the MTLCNN performs strongly, with most final $R^{2}$ errors above $0.99$. The MTLCNN performs worse on the test dataset at lower bulk Reynolds numbers (Table~\ref{curriculum_learning_std_training_table}) because the flow is at its marginal state and at the same, we are also at the edge of the sweeping parameter space. Note that although in Case 1 and Case 2 the flow is fully-developed, there is still a drop in the $R^{2}$ performance for the test dataset since those two cases are at the edge of the sweeping parameters space too. Nevertheless, the $R^{2}$ does not decrease even $2\%$ compared to other cases. Figures~\ref{fig:prediction1350uv}, displays plots of the component predictions for Case 12 alongside their DNS labels. The predictions compare favorably to the true labels: they successfully capture the main trends and the behavior present in the real data, with only minor discrepancies. For all other bulk Reynolds numbers, the same is true.

\begin{table}[htb!]
\caption{$R^{2}$ error for the trained MTLCNN model predictions. The results before filtering and boosting using standard training and curriculum training are shown on the first two columns. The column on the right displays the final model results after applying filtering and boosting (see Section~\ref{subsec: Filtering and Boosting}) to the prediction obtained using curriculum learning. Curriculum learning does in general boost the performance. Nevertheless, standard training does already give a good performance.}
\centering
\label{curriculum_learning_std_training_table}
\resizebox{0.9\columnwidth}{!}{\begin{tabular}{ccccc}\toprule
\multicolumn{1}{l}{Case} & \multicolumn{1}{l}{} & \multicolumn{1}{l}{Standard Training} & \multicolumn{1}{l}{Curriculum Training}& \multicolumn{1}{l}{After Filtering and Boosting} \\\midrule
\multirow{2}{*}{1}       & Train                & 0.9902                                & 0.9937     & \textbf{0.9954}                             \\
                         & Test                 & 0.9805                                & 0.9898  & \textbf{0.9903}                                \\\midrule
\multirow{2}{*}{2}       & Train                & 0.9877                                & 0.9918    & \textbf{0.9942}                              \\
                         & Test                 & 0.9896                                & 0.9917     & \textbf{0.9922}                             \\\midrule
\multirow{2}{*}{3}       & Train                & 0.9896                                & 0.9928  & \textbf{0.9948}                                 \\
                         & Test                 & 0.9912                                & 0.9930   & \textbf{0.9943}                                \\\midrule
\multirow{2}{*}{4}       & Train                & 0.9907                                & 0.9927        & \textbf{0.9947}                           \\
                         & Test                 & 0.9892                                & 0.9931      & \textbf{0.9953}                             \\\midrule
\multirow{2}{*}{5}       & Train                & 0.9931                                & 0.9940       & \textbf{0.9953}                            \\
                         & Test                 & 0.9928                                & 0.9955   & \textbf{0.9967}                                \\\midrule
\multirow{2}{*}{6}       & Train                & 0.9920                                & 0.9920       & \textbf{0.9940}                            \\
                         & Test                 & 0.9924                                & 0.9925    & \textbf{0.9955}                               \\\midrule
\multirow{2}{*}{7}       & Train                & 0.9923                                & 0.9921     & \textbf{0.9940}                              \\
                         & Test                 & 0.9901                                & 0.9923  & \textbf{0.9962}                                 \\\midrule
\multirow{2}{*}{8}       & Train                & 0.9930                                & 0.9936     & \textbf{0.9944}                             \\
                         & Test                 & 0.9950                                & 0.9952  & \textbf{0.9976}                                 \\\midrule
\multirow{2}{*}{9}       & Train                & 0.9913                                & 0.9914     & \textbf{0.9943}                              \\
                         & Test                 & 0.9934                                & 0.9936  & \textbf{0.9971}                                 \\\midrule
\multirow{2}{*}{10}      & Train                & 0.9910                                & 0.9923 & \textbf{0.9939}                                  \\
                         & Test                 & 0.9934                                & 0.9955    & \textbf{0.9966}                               \\\midrule
\multirow{2}{*}{11}      & Train                & 0.9894                                & 0.9922    & \textbf{0.9940}                               \\
                         & Test                 & 0.9896                                & 0.9897   & \textbf{0.9921}                                \\\midrule
\multirow{2}{*}{12}      & Train                & 0.9899                                & 0.9926  & \textbf{0.9942}                                 \\
                         & Test                 & 0.9894                                & 0.9950     & \textbf{0.9974}                              \\\midrule
\multirow{2}{*}{13}      & Train                & 0.9921                                & 0.9936  & \textbf{0.9952}                                 \\
                         & Test                 & 0.9870                                & 0.9870   & \textbf{0.9889}                                \\\midrule
\multirow{2}{*}{14}      & Train                & 0.9919                                & 0.9921    & \textbf{0.9940}                               \\
                         & Test                 & 0.9875                                & 0.9886   & \textbf{0.9925}                                \\\midrule
\multirow{2}{*}{15}      & Train                & 0.9917                                & 0.9927       & \textbf{0.9944}                            \\
                         & Test                 & 0.9795                                & 0.9818    & \textbf{0.9811}                               \\\midrule
\multirow{2}{*}{16}      & Train                & 0.9916                                & 0.9920    & \textbf{0.9944}                               \\
                         & Test                 & 0.9773                                & 0.9838   & \textbf{0.9800}  \\\bottomrule                             
\end{tabular}}
\end{table}

\clearpage

\clearpage
\section{Conclusion}
\label{sec:Conclusion}

Modeling the Reynolds stress tensor is a central, challenging problem in turbulence modeling. Researchers have started to apply ML algorithms to tackle this problem. In this work we presented a novel deep convolutional neural network based on multi-task learning for turbulent duct flow. The new MTLCNN model is an extension of the convolutional models proposed in~\cite{borde2021convolutional} for turbulent one-dimensional flow to turbulent duct flow. Additionally, we also explored curriculum learning and the findings suggest that this technique could help speed up the convergence and improve the performance of deep data-driven turbulence models. Overall, we obtain excellent results across all bulk Reynolds numbers in the dataset using a network with very few training parameters given the complexity of the problem.

An accurate prediction of the Reynolds stress tensor does not necessarily imply accuracy in the mean velocity field~\cite{duraisamy}. In the work by~\cite{thompson_errors_dns} a number of databases were tested to verify if the available Reynolds stress tensor could recover the mean velocity field, and large propagation errors were found, especially for high Reynolds numbers. This issue was also studied by~\cite{wu_ill_conditioned} and~\cite{brener_cruz_thompson_anjos_2021}.

There are several directions for future work. Developing new convolutional models able to deal with complicated boundary geometries and unstructured meshes would expand the applicability of the networks to a great number of practical flows. Exploring transfer learning between high-dimensional turbulent flows and trying similar arquitectures as the MTLCNN for external flows are also interesting avenues for future research.

\newpage
\bibliography{biblio}

\begin{thebibliography}{}

\bibitem[Allgower and Georg, 2003]{continuation}
Allgower, E. and Georg, K. (2003).
\newblock {\em Numerical Continuation Methods—An Introduction}, volume xxv.

\bibitem[Aloui et~al., 2011]{agrofood}
Aloui, F., Berrich, E., and Pierrat, D. (2011).
\newblock {Experimental and Numerical Investigations of a Turbulent Flow
  Behavior in Isolated and Nonisolated Conical Diffusers}.
\newblock {\em Journal of Fluids Engineering}, 133(1).
\newblock 011201.

\bibitem[Aly, 2005]{Aly2005SurveyOM}
Aly, M. (2005).
\newblock Survey on multiclass classification methods.

\bibitem[Anxionnaz-Minvielle et~al., 2013]{process}
Anxionnaz-Minvielle, Z., Cabassud, M., Gourdon, C., and Tochon, P. (2013).
\newblock Influence of the meandering channel geometry on the thermo-hydraulic
  performances of an intensified heat exchanger/reactor.
\newblock {\em Chemical Engineering and Processing: Process Intensification},
  73:67–80.

\bibitem[Baxter, 1997]{Baxter97abayesian/information}
Baxter, J. (1997).
\newblock A bayesian/information theoretic model of learning to learn via
  multiple task sampling.
\newblock In {\em Machine Learning}, pages 7--39.

\bibitem[Baxter, 2011]{bias_learning}
Baxter, J. (2011).
\newblock A model of inductive bias learning.
\newblock {\em CoRR}, abs/1106.0245.

\bibitem[Bengio et~al., 2009]{cl_bengio}
Bengio, Y., Louradour, J., Collobert, R., and Weston, J. (2009).
\newblock Curriculum learning.
\newblock volume~60, page~6.

\bibitem[Boscaini et~al., 2016]{NIPS2016_228499b5}
Boscaini, D., Masci, J., Rodol\`{a}, E., and Bronstein, M. (2016).
\newblock Learning shape correspondence with anisotropic convolutional neural
  networks.
\newblock In Lee, D., Sugiyama, M., Luxburg, U., Guyon, I., and Garnett, R.,
  editors, {\em Advances in Neural Information Processing Systems}, volume~29.
  Curran Associates, Inc.

\bibitem[Bottou et~al., 2016]{ml_opt}
Bottou, L., Curtis, F.~E., and Nocedal, J. (2016).
\newblock Optimization methods for large-scale machine learning.
\newblock {\em SIAM Review}, 60.

\bibitem[Bradshaw, 2003]{peter}
Bradshaw, P. (2003).
\newblock Turbulent secondary flows.
\newblock {\em Annual Review of Fluid Mechanics}, 19:53--74.

\bibitem[Brener et~al., 2021]{brener_cruz_thompson_anjos_2021}
Brener, B.~P., Cruz, M.~A., Thompson, R.~L., and Anjos, R.~P. (2021).
\newblock Conditioning and accurate solutions of reynolds average
  navier–stokes equations with data-driven turbulence closures.
\newblock {\em Journal of Fluid Mechanics}, 915:A110.

\bibitem[Brundrett and Baines, 1964]{brundrett_baines_1964}
Brundrett, E. and Baines, W.~D. (1964).
\newblock The production and diffusion of vorticity in duct flow.
\newblock {\em Journal of Fluid Mechanics}, 19(3):375–394.

\bibitem[Chang and Tavoularis, 2007]{nuclear}
Chang, D.-i. and Tavoularis, S. (2007).
\newblock Numerical simulation of turbulent flow in a 37-rod bundle.
\newblock {\em Nuclear Engineering and Design}, 237:575--590.

\bibitem[Chen et~al., 2003]{chen2003extended}
Chen, H., Kandasamy, S., Orszag, S., Shock, R., Succi, S., and Yakhot, V.
  (2003).
\newblock Extended boltzmann kinetic equation for turbulent flows.
\newblock {\em Science}, 301(5633):633--636.

\bibitem[Collobert and Weston, 2008]{Collobert2008AUA}
Collobert, R. and Weston, J. (2008).
\newblock A unified architecture for natural language processing: deep neural
  networks with multitask learning.
\newblock In {\em ICML '08}.

\bibitem[Craft et~al., 1996]{craft}
Craft, T., Launder, B., and Suga, K. (1996).
\newblock Development and application of a cubic eddy-viscosity model of
  turbulence.
\newblock {\em International Journal of Heat and Fluid Flow}, 17:108--115.

\bibitem[Deng et~al., 2013]{Deng2013NewTO}
Deng, L., Hinton, G.~E., and Kingsbury, B. (2013).
\newblock New types of deep neural network learning for speech recognition and
  related applications: an overview.
\newblock {\em 2013 IEEE International Conference on Acoustics, Speech and
  Signal Processing}, pages 8599--8603.

\bibitem[Duong et~al., 2015]{Duong2015LowRD}
Duong, L., Cohn, T., Bird, S., and Cook, P. (2015).
\newblock Low resource dependency parsing: Cross-lingual parameter sharing in a
  neural network parser.
\newblock In {\em ACL}.

\bibitem[Duraisamy et~al., 2019]{duraisamy}
Duraisamy, K., Iaccarino, G., and Xiao, H. (2019).
\newblock Turbulence modeling in the age of data.
\newblock {\em Annual Review of Fluid Mechanics}, 51(1):357--377.

\bibitem[Durbin, 1991]{Durbin1991NearwallTC}
Durbin, P. (1991).
\newblock Near-wall turbulence closure modeling without “damping
  functions”.
\newblock {\em Theoretical and Computational Fluid Dynamics}, 3:1--13.

\bibitem[Elman, 1993]{ELMAN199371}
Elman, J.~L. (1993).
\newblock Learning and development in neural networks: the importance of
  starting small.
\newblock {\em Cognition}, 48(1):71--99.

\bibitem[Fang et~al., 2018]{fang2018deep}
Fang, R., Sondak, D., Protopapas, P., and Succi, S. (2018).
\newblock Deep learning for turbulent channel flow.

\bibitem[Gatski, 2004]{gatski2004constitutive}
Gatski, T. (2004).
\newblock Constitutive equations for turbulent flows.
\newblock {\em Theoretical and Computational Fluid Dynamics}, 18(5):345--369.

\bibitem[Gerolymos et~al., 2010]{duct_aerospace}
Gerolymos, G., Joly, S., Mallet, M., and Vallet, I. (2010).
\newblock Reynolds-stress model flow prediction in aircraft-engine intake
  double-s-shaped duct.
\newblock {\em Journal of Aircraft - J AIRCRAFT}, 47:1368--1381.

\bibitem[Gerolymos et~al., 2004]{gerorms}
Gerolymos, G., Sauret, E., and Vallet, I. (2004).
\newblock Contribution to the single-point-closure reynolds–stress modelling
  of inhomogeneous flow.
\newblock {\em Theoretical and Computational Fluid Dynamics}, 17:407--431.

\bibitem[Gerolymos and Vallet, 2015]{Gerolymos_2015}
Gerolymos, G.~A. and Vallet, I. (2015).
\newblock Reynolds-stress model prediction of 3-d duct flows.
\newblock {\em Flow, Turbulence and Combustion}, 96(1):45–93.

\bibitem[Gessner and Jones, 1965]{gessner_jones_1965}
Gessner, F.~B. and Jones, J.~B. (1965).
\newblock On some aspects of fully-developed turbulent flow in rectangular
  channels.
\newblock {\em Journal of Fluid Mechanics}, 23(4):689–713.

\bibitem[Ghosn and Bengio, 1996]{stock_mt}
Ghosn, J. and Bengio, Y. (1996).
\newblock Multi-task learning for stock selection.
\newblock pages 946--952.

\bibitem[Girshick, 2015]{fast_r_cnn}
Girshick, R. (2015).
\newblock Fast r-cnn.
\newblock In {\em Proceedings of the 2015 IEEE International Conference on
  Computer Vision (ICCV)}, ICCV '15, page 1440–1448, USA. IEEE Computer
  Society.

\bibitem[Goodfellow et~al., 2016]{Goodfellow-et-al-2016}
Goodfellow, I., Bengio, Y., and Courville, A. (2016).
\newblock {\em Deep Learning}.
\newblock MIT Press.
\newblock \url{http://www.deeplearningbook.org}.

\bibitem[He et~al., 2015]{he2015delving}
He, K., Zhang, X., Ren, S., and Sun, J. (2015).
\newblock Delving deep into rectifiers: Surpassing human-level performance on
  imagenet classification.
\newblock In {\em Proceedings of the IEEE international conference on computer
  vision}, pages 1026--1034.

\bibitem[Jiang et~al., 2021]{jiang}
Jiang, C., Vinuesa, R., Chen, R., Mi, J., Laima, S., and Li, H. (2021).
\newblock An interpretable framework of data-driven turbulence modeling using
  deep neural networks.
\newblock {\em Physics of Fluids}, 33:055133.

\bibitem[Johansson, 2002]{johansson2002engineering}
Johansson, A. (2002).
\newblock Engineering turbulence models and their development, with emphasis on
  explicit algebraic reynolds stress models.
\newblock In {\em Theories of Turbulence}, pages 253--300. Springer.

\bibitem[Kaandorp, 2018]{Kaandorp2018MachineLF}
Kaandorp, M. (2018).
\newblock Machine learning for data-driven rans turbulence modelling.

\bibitem[Kaandorp and Dwight, 2020]{KAANDORP2020104497}
Kaandorp, M.~L. and Dwight, R.~P. (2020).
\newblock Data-driven modelling of the reynolds stress tensor using random
  forests with invariance.
\newblock {\em Computers \& Fluids}, 202:104497.

\bibitem[Kiefer and Wolfowitz, 1952]{sgd}
Kiefer, J. and Wolfowitz, J. (1952).
\newblock {Stochastic Estimation of the Maximum of a Regression Function}.
\newblock {\em The Annals of Mathematical Statistics}, 23(3):462 -- 466.

\bibitem[Lam, 1983]{interpolation}
Lam, N. (1983).
\newblock Spatial interpolation methods: a review.
\newblock {\em American Cartographer}, 10:129--149.

\bibitem[Lee and Moser, 2018]{dns_couette}
Lee, M. and Moser, R. (2018).
\newblock Extreme-scale motions in turbulent plane couette flows.
\newblock {\em Journal of Fluid Mechanics}, 842:128--145.

\bibitem[Lien and Kalitzin, 2001]{lien_version}
Lien, F.-S. and Kalitzin, G. (2001).
\newblock Computations of transonic flow with the v2–f turbulence model.
\newblock {\em International Journal of Heat and Fluid Flow}, 22:53--61.

\bibitem[Ling et~al., 2016a]{ling_ml}
Ling, J., Jones, R., and Templeton, J. (2016a).
\newblock Machine learning strategies for systems with invariance properties.
\newblock {\em Journal of Computational Physics}, 318.

\bibitem[Ling et~al., 2016b]{ling2016reynolds}
Ling, J., Kurzawski, A., and Templeton, J. (2016b).
\newblock Reynolds averaged turbulence modelling using deep neural networks
  with embedded invariance.
\newblock {\em Journal of Fluid Mechanics}, 807:155--166.

\bibitem[Marcos et~al., 2016]{GonzalezVT16}
Marcos, D., Volpi, M., and Tuia, D. (2016).
\newblock Learning rotation invariant convolutional filters for texture
  classification.
\newblock {\em CoRR}, abs/1604.06720.

\bibitem[Masci et~al., 2015]{MasciBBV15}
Masci, J., Boscaini, D., Bronstein, M.~M., and Vandergheynst, P. (2015).
\newblock Shapenet: Convolutional neural networks on non-euclidean manifolds.
\newblock {\em CoRR}, abs/1501.06297.

\bibitem[McConkey et~al., 2021]{mcconkey2021curated}
McConkey, R., Yee, E., and Lien, F.-S. (2021).
\newblock A curated dataset for data-driven turbulence modelling.

\bibitem[Millstein, 2018]{millstein2018convolutional}
Millstein, F. (2018).
\newblock {\em Convolutional Neural Networks in Python: Beginner's Guide to
  Convolutional Neural Networks in Python}.
\newblock CreateSpace Independent Publishing Platform.

\bibitem[Moin and Mahesh, 1998]{dns_review}
Moin, P. and Mahesh, K. (1998).
\newblock Direct numerical simulation: A tool in turbulence research.
\newblock {\em Annual Review of Fluid Mechanics}, 30(1):539--578.

\bibitem[Monti et~al., 2016]{monti}
Monti, F., Boscaini, D., Masci, J., Rodol{\`{a}}, E., Svoboda, J., and
  Bronstein, M.~M. (2016).
\newblock Geometric deep learning on graphs and manifolds using mixture model
  cnns.
\newblock {\em CoRR}, abs/1611.08402.

\bibitem[Nikitin et~al., 2019]{secondary}
Nikitin, N., Pimanov, V., and Popelenskaya, N. (2019).
\newblock Mechanism of formation of prandtl’s secondary flows of the second
  kind.
\newblock {\em Doklady Physics}, 484:420--425.

\bibitem[Pan and Yang, 2010]{5288526}
Pan, S.~J. and Yang, Q. (2010).
\newblock A survey on transfer learning.
\newblock {\em IEEE Transactions on Knowledge and Data Engineering},
  22(10):1345--1359.

\bibitem[Pinelli et~al., 2010]{pinelli_uhlmann_sekimoto_kawahara_2010}
Pinelli, A., Uhlmann, M., Sekimoto, A., and Kawahara, G. (2010).
\newblock Reynolds number dependence of mean flow structure in square duct
  turbulence.
\newblock {\em Journal of Fluid Mechanics}, 644:107–122.

\bibitem[Pope, 1975]{pope_1975}
Pope, S.~B. (1975).
\newblock A more general effective-viscosity hypothesis.
\newblock {\em Journal of Fluid Mechanics}, 72(2):331–340.

\bibitem[Pope, 2000]{alma993553114401591}
Pope, S.~B. (2000).
\newblock {\em Turbulent flows}.
\newblock Cambridge University Press, Cambridge.

\bibitem[Poulenard and Ovsjanikov, 2018]{Poulenard2018MultidirectionalGN}
Poulenard, A. and Ovsjanikov, M. (2018).
\newblock Multi-directional geodesic neural networks via equivariant
  convolution.
\newblock {\em ACM Transactions on Graphics (TOG)}, 37:1 -- 14.

\bibitem[Ramsundar et~al., 2015]{ramsundar2015massively}
Ramsundar, B., Kearnes, S., Riley, P., Webster, D., Konerding, D., and Pande,
  V. (2015).
\newblock Massively multitask networks for drug discovery.

\bibitem[Rohde and Plaut, 1999]{rohde}
Rohde, D. and Plaut, D. (1999).
\newblock Language acquisition in the absence of explicit negative evidence:
  how important is starting small?
\newblock {\em Cognition}, 72(1):67—109.

\bibitem[Ruder, 2016]{DBLP:journals/corr/Ruder16}
Ruder, S. (2016).
\newblock An overview of gradient descent optimization algorithms.
\newblock {\em CoRR}, abs/1609.04747.

\bibitem[So and Yuan, 1999]{sormc}
So, R. and Yuan, S. (1999).
\newblock A geometry independent near-wall reynolds-stress closure.
\newblock {\em International Journal of Engineering Science}, Vol. 37:33--57.

\bibitem[Song et~al., 2019]{song}
Song, X., Zhang, Z., Wang, Y., Ye, S., and Huang, C. (2019).
\newblock Reconstruction of rans model and cross-validation of flow field based
  on tensor basis neural network.

\bibitem[Speziale, 1991]{speziale}
Speziale, C.~G. (1991).
\newblock Analytical methods for the development of reynolds-stress closures in
  turbulence.
\newblock {\em Annual Review of Fluid Mechanics}, 23(1):107--157.

\bibitem[Sáez~de Ocáriz~Borde et~al., 2021]{borde2021convolutional}
Sáez~de Ocáriz~Borde, H., Sondak, D., and Protopapas, P. (2021).
\newblock Convolutional neural network models and interpretability for the
  anisotropic reynolds stress tensor in turbulent one-dimensional flows.

\bibitem[Thompson et~al., 2016]{thompson_errors_dns}
Thompson, R., Sampaio, L., de~Bragança~Alves, F., Thais, L., and Mompean, G.
  (2016).
\newblock A methodology to evaluate statistical errors in dns data of plane
  channel flows.
\newblock {\em Computers $\&$ Fluids}, 130.

\bibitem[Thung and Wee, 2018]{mt_thung}
Thung, K. and Wee, C.-Y. (2018).
\newblock A brief review on multi-task learning.
\newblock {\em Multimedia Tools and Applications}, 77.

\bibitem[Uhlmann et~al., 2007]{marginal}
Uhlmann, M., Pinelli, A., KAWAHARA, G., and Sekimoto, A. (2007).
\newblock Marginally turbulent flow in a square duct.
\newblock {\em Journal of Fluid Mechanics}, 588:153 -- 162.

\bibitem[Vafaeikia et~al., 2020]{multi-task_learning}
Vafaeikia, P., Namdar, K., and Khalvati, F. (2020).
\newblock A brief review of deep multi-task learning and auxiliary task
  learning.
\newblock {\em CoRR}, abs/2007.01126.

\bibitem[Vallet, 2007]{3dre}
Vallet, I. (2007).
\newblock {Reynolds-Stress Modeling of Three-Dimensional Secondary Flows With
  Emphasis on Turbulent Diffusion Closure}.
\newblock {\em Journal of Applied Mechanics}, 74(6):1142--1156.

\bibitem[Vinuesa et~al., 2018]{ducts_vinuesa}
Vinuesa, R., Schlatter, P., and Nagib, H. (2018).
\newblock Secondary flow in turbulent ducts with increasing aspect ratio.
\newblock {\em Physical Review Fluids}, 3:054606.

\bibitem[Wang et~al., 2020]{cl_survey}
Wang, X., Chen, Y., and Zhu, W. (2020).
\newblock A comprehensive survey on curriculum learning.
\newblock {\em CoRR}, abs/2010.13166.

\bibitem[Wu et~al., 2017]{laizet}
Wu, J., Sun, R., Laizet, S., and Xiao, H. (2017).
\newblock Representation of reynolds stress perturbations with application in
  machine-learning-assisted turbulence modeling.
\newblock {\em Computer Methods in Applied Mechanics and Engineering}, 346.

\bibitem[Wu et~al., 2016]{wu_again}
Wu, J., Wang, J.-X., and Xiao, H. (2016).
\newblock A bayesian calibration-prediction method for reducing model-form
  uncertainties with application in rans simulations.
\newblock {\em Flow, Turbulence and Combustion}, 97.

\bibitem[Wu et~al., 2019]{wu_ill_conditioned}
Wu, J., Xiao, H., Sun, R., and Wang, Q. (2019).
\newblock Rans equations with explicit data-driven reynolds stress closure can
  be ill-conditioned.

\bibitem[Wu et~al., 2018]{PhysRevFluids.3.074602}
Wu, J.-L., Xiao, H., and Paterson, E. (2018).
\newblock Physics-informed machine learning approach for augmenting turbulence
  models: A comprehensive framework.
\newblock {\em Phys. Rev. Fluids}, 3:074602.

\bibitem[Wu et~al., 2020]{senwu}
Wu, S., Zhang, H.~R., and R{\'{e}}, C. (2020).
\newblock Understanding and improving information transfer in multi-task
  learning.
\newblock {\em CoRR}, abs/2005.00944.

\bibitem[Yang and Hospedales, 2017]{Yang2017TraceNR}
Yang, Y. and Hospedales, T.~M. (2017).
\newblock Trace norm regularised deep multi-task learning.
\newblock {\em ArXiv}, abs/1606.04038.

\bibitem[Zhang and Yang, 2021]{zhang2021surveymt}
Zhang, Y. and Yang, Q. (2021).
\newblock A survey on multi-task learning.

\bibitem[Zhu and Dinh, 2020]{zhu2020datadriven}
Zhu, Y. and Dinh, N. (2020).
\newblock A data-driven approach for turbulence modeling.

\end{thebibliography}
\bibliographystyle{apalike}

\end{document}